\documentclass[twocolumn,notitlepage,pra,superscriptaddress,longbibliography]{revtex4-1}

\usepackage{textcomp}
\usepackage{times}
\usepackage{graphicx}
\usepackage{float}
\usepackage{latexsym,amsmath,amssymb,bm,euscript}
\usepackage{xcolor}
\usepackage{subfigure}
\usepackage{epstopdf}
\usepackage[breaklinks=true,colorlinks=true,linkcolor=blue,urlcolor=blue,citecolor=blue]{hyperref}
\usepackage{hyperref}
\usepackage{soul}
\usepackage[normalem]{ulem}
\usepackage{mathrsfs}
\usepackage{amsmath}
\usepackage{xspace}
\usepackage{CJKutf8}
\usepackage{placeins}
\usepackage{appendix}

\graphicspath{{./Figs/}}

\begin{document}
%\begin{CJK}{UTF8}{<font>}
\title{Multitype entanglement dynamics induced by exceptional points}

\author{Zigeng Li}
\affiliation{School of Physics, Beihang University, Beijing 100191, China}

\author{Xinyao Huang}
\email{xinyaohuang@buaa.edu.cn}
\affiliation{School of Physics, Beihang University, Beijing 100191, China}

\author{Hongyan Zhu}
\affiliation{School of Physics, Beihang University, Beijing 100191, China}

\author{Guofeng Zhang}
\email{gf1978zhang@buaa.edu.cn}
\affiliation{School of Physics, Beihang University, Beijing 100191, China}

\author{Fan Wang}
\affiliation{School of Physics, Beihang University, Beijing 100191, China}

\author{Xiaolan Zhong}
\email{zhongxl@buaa.edu.cn}
\affiliation{School of Physics, Beihang University, Beijing 100191, China}

\begin{abstract}
    As a most important feature of non-Hermitian systems, exceptional points (EPs) lead to a variety of unconventional phenomena and applications. 
    %However, the majority of studies focused on the EP-induced classical effects.
    Here, we study a generic model composed of two coupled non-Hermitian qubits, the EPs can be easily obtained in this system by adjusting the driving amplitude applied to the qubits. The diverse entanglement dynamics on the two sides of the original fourth-order EP (EP4') and second order EP (EP2) can be observed simultaneously in the weak coupling regime. With the increase of the coupling strength, the EP4' is replaced by an additional EP2, leading to the disappearance of the entanglement dynamics changing induced by EP4' in the strong coupling regime.
    Considering the case of Ising type interaction, we also realize EP-induced entanglement dynamics changing without the driving field.
    Our study paves the way for the investigation of EP-induced quantum effects and applications of EP-related quantum technologies.
    \end{abstract}
\date{\today} % Wb,Jan18,19
\maketitle

\section{Introduction}
\label{1}
Non-Hermitian (NH) Hamiltonians provide an effective method to describe physical systems that exchange energy, information or particles with the environment \cite{PhysRevLett.80.5243, RN454}. %加参考文献
For instance, NH Hamiltonians have been applied to describe optical \cite{RN475,RN476,RN479}, acoustic \cite{PhysRevLett.121.124501,PhysRevLett.120.124502} and magnetic systems \cite{PhysRevLett.125.147202,PhysRevLett.123.127202} with gain and loss or asymmetric couplings.
These non-Hermitian systems have unprecedented features induced by non-Hermiticity\cite{PhysRevLett.80.5243}.
As a most peculiar example, exceptional points (EPs) describe the coalescence of the eigenstates and the degeneracy of the eigenvalues, indicating the phase transition from the parity-time ($\mathcal{PT}$) symmetric phase to the $\mathcal{PT}$-symmetry-broken phase \cite{RN454,RevModPhys.93.015005,RN494,RN490,RN491}. 
EPs play an important role in many remarkable physical phenomena and functional applications. For example, the nonlinear perturbation response of EPs induces a variety of unconventional effects, including nonreciprocal light propagation \cite{RN485,RN488}, loss-induced transparency \cite{RN467,RN470} and  coherent perfect absorption \cite{PhysRevLett.112.143903,RN480}, making them promising platforms for wireless power transfer \cite{RN483,RN484}, single-mode lasers \cite{RN466} and so on.%现象和功能分开举例，再补充一些

In addition to investigating EPs in classical systems, the extension of EPs to quantum regime has attracted much attention.
%验证一下这些文献的方法是不是都可以归为这两类
Based on the methods of extended Hilbert space \cite{RN460} and quantum trajectories \cite{RN461} to construct effective NH Hamiltonians, EPs have been achieved in various quantum platforms such as dissipative photon systems \cite{RN462,RN463}, superconducting circuits \cite{RN461,PhysRevLett.127.140504,RN472,PhysRevLett.128.110402}, single ion traps \cite{PhysRevLett.126.083604}, thermal atom ensembles \cite{PhysRevLett.124.030401,PhysRevLett.130.263601}, cold atoms \cite{RN498,PhysRevX.4.041001} and nitrogen-vacancy color-centers \cite{RN460,PhysRevResearch.1.013015}. Quantum effects including photon blockade\cite{RN496,RN497,PhysRevA.106.043715,PhysRevB.108.134409} and topological quantum state control\cite{PhysRevLett.126.170506,PhysRevX.8.031079,RN471} have been achieved by engineering the EPs. 
As a most important quantum property, EP-induced exceptional entanglement behaviors have also been observed in very recent works. For example, the occurrence of the entanglement transition at the EP has been observed in a NH system composed of qubit coupled with photons\cite{PhysRevLett.131.260201}. Entanglement generation between two NH qubits can be accelerated by approaching the EP of the system\cite{PhysRevLett.131.100202}. Replacing one NH qubit with unitary qubit, entanglement can be maximized at the EP\cite{PhysRevA.105.012422}. 
However, the relationship between the EPs and entanglement dynamics remains unknown.

Here, we show that multitype entanglement dynamics can be realized by engineering the EPs, by analyzing a generic model composed of two coupled NH qubits. The inter-qubit coupling will lower the order of EP from the original fourth-order EP (EP4) to a second-order EP (EP2). However, we find that disparate entanglement dynamics on the two sides of the original EP4 and EP2 can be observed simultaneously in the case of weak coupling (i.e., the coupling strength is much smaller than the dissipation rate), indicating different types of entanglement behaviors can be induced by the EPs. When enhancing the coupling strength to the strong coupling regime (i.e., the coupling strength and dissipation rate are on the same order), the  original EP4 is replaced by an additional EP2, leading to the disappearance of the EP4-engineered entanglement changing. Taking Ising type interaction as an example, we also find EP-induced entanglement changing in the absence of the driving field. 
Our scheme has good experimental feasibility, different types of platforms can be used to study the simulation of non-Hermitian systems and the observation of EPs, such as the coupled loss and gain cavities \cite{RN663,RN669}, cavity optomechanical systems \cite{RN665, RN666}, superconducting circuits \cite{RN461,RN499,RevModPhys.85.623,RN500}, ion traps \cite{RN501,RN502} and single-spin system \cite{RN460}. We believe that our study provides an efficient method for designing EP based quantum devices, such as the quantum-state synthesis using exceptional-point engineering in superconducting resonators \cite{PhysRevResearch.5.033119}, single-photon interferometric network \cite{RN507,PhysRevLett.123.230401} and enhancing the performance of quantum metrology near the EP \cite{PhysRevLett.117.110802}. 

The paper is organized as follows. Section \ref{2} presents theoretical models for the two coupled non-Hermitian qubits. In Section \ref{3}, we conduct numerical simulations for predicting the EPs in the system. We show the entanglement dynamics change in Section \ref{4} and discuss the reason of this phenomenon in Section \ref{5}. Furthermore, we demonstrate that the entanglement dynamics persist in systems with dipolar coupling, as discussed in Section \ref{6} and \ref{7}, respectively.  We summarize our findings in Section \ref{8}.

\begin{figure}[!tbp]
\includegraphics[angle=0,width=\linewidth]{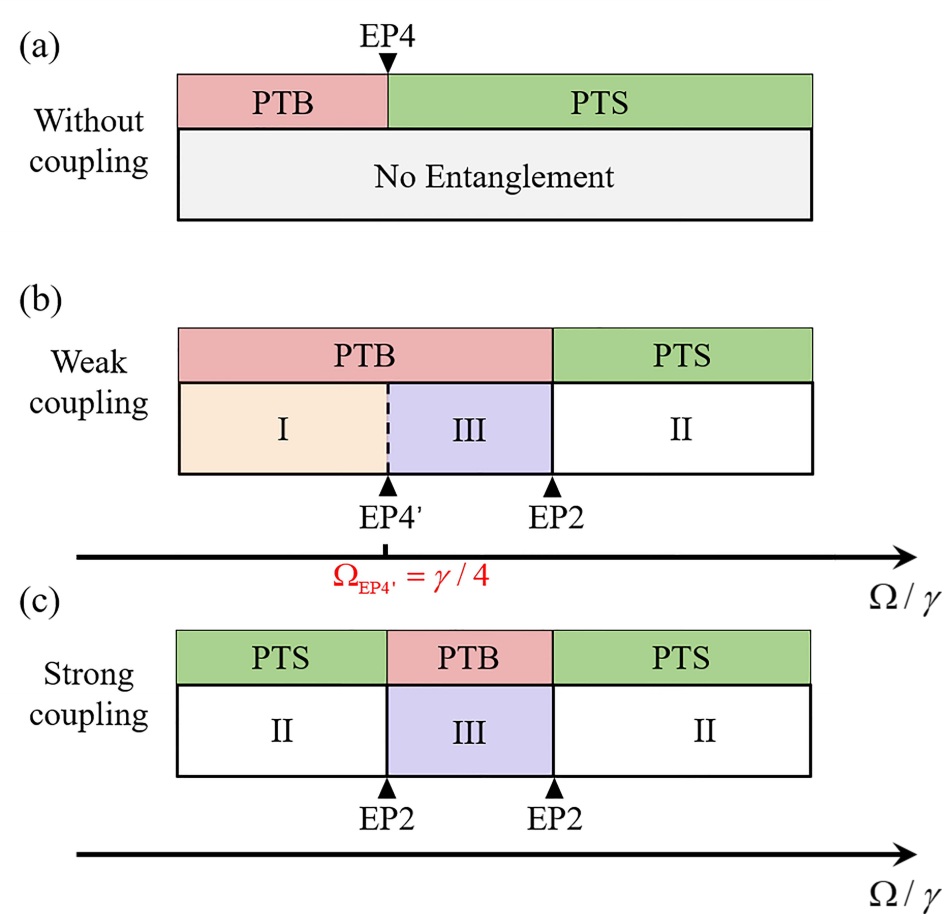}
\caption{
Schematic diagram of the multitype entanglement dynamics (I, II, III) induced by EPs with increasing $\Omega$. The regions of PTB and PTS represent the $\mathcal{P}\mathcal{T}$-symmetry-broken phase and the $\mathcal{P}\mathcal{T}$-symmetric phase, respectively. The EPs can be flexibly changed by tuning the drive amplitude $\Omega$. (a) The system exhibits a fourth-order EP (EP4) without coupling and no entanglement will be generated in this case. (b) In the weak coupling regime (the coupling strength $\xi$ is much smaller than the dissipation rate $\gamma$ of the qubits, i.e., $\xi \ll \gamma$), the system exhibits the original EP4 (EP4') and an existing second-order EP (EP2). (c) In the strong coupling regime (i.e, $\xi \sim\gamma $), the original EP4 disappears and an additional EP2 emerges. Corresponding to the presence of EP4', the purely imaginary (PI) eigenvalues become a mixture of both real and imaginary parts. The EP2 induces the spectrum transition of the PTS and the mixed phase. The orange, white and blue represent the region of the PI, PTS and mixed phase, respectively. The entanglement behavior of type I in PI is monotonically increasing and eventually reaching a stable value, type II in PTS corresponds to the entanglement behavior of continuous oscillations over the entire time domain, and type III in the mixed phase combines the characteristics of the above two different entanglement behaviors.
}
\label{Fig:1}
\end{figure}

\section{Model}
\label{2}

We consider a generic system composed of two coupled NH qubits (or spin). The type of interaction can be treated as either dipolar or Ising type interaction.
The system Hamiltonian is given as ($\hbar =1$)
\begin{equation}
{{H}_{j=1,2}}=\sum\limits_{j=1,2}{\left( -\frac{{{\gamma }_{j}}}{2}\sigma _{j}^{-}\sigma _{j}^{+}+{{\Omega }_{j}}\sigma _{j}^{x} \right)+{{H}_{\operatorname{int}}}}.
\label{Eq:1}
\end{equation}
The Pauli operators in terms of the quantum energy levels-with no classical analogs- ${{\left| b \right\rangle }_{j}}$ and ${{\left| a \right\rangle }_{j}}$ as $\sigma _{j}^{+}={{\left| a \right\rangle }_{j}}\left\langle  b \right|, \sigma _{j}^{-}={{\left| b \right\rangle }_{j}}\left\langle  a \right|$ and $\sigma _{j}^{x}={{\left| b \right\rangle }_{j}}\left\langle  a \right|+{{\left| a \right\rangle }_{j}}\left\langle  b \right|,j\in\{1,2\}$. ${{\gamma }_{j}}$  and ${{\Omega }_{j}}$ \textcolor{red}{are} the energy decay rate of ${{\left| b \right\rangle }_{j}}$ and driving amplitude, respectively. 
In the following analysis, we assume resonance between the driving field frequency and the qubit transition, with ${{\gamma }_{1}}={{\gamma }_{2}}=\gamma $ and ${{\Omega }_{1}}={{\Omega }_{2}}=\Omega $ for simplicity.
The third term represents the  interaction between the two NH qubits. When the two qubits are coupled via dipolar interaction, the Hamiltonian can be described as ${{H}_{\operatorname{int}}}=g\left( \sigma _{1}^{\dagger }{{\sigma }_{2}}+{{\sigma }_{1}}\sigma _{2}^{\dagger } \right)$, which can be realized in various systems such as superconducting circuit systems \cite{RN461,PhysRevB.92.174507} and Rydberg atom systems \cite{RN482}.
In addition to this interaction type, there is another type of interaction between the two  qubits or spins, i.e., Ising type interaction, which can be represented as ${{H}_{\operatorname{int}}}=-\xi {{\left( \sigma _{1}^{x}+\sigma _{2}^{x} \right)}^{2}}$. Here $g$ and $\xi $ denote the coupling strength, respectively. Ising type interaction has been widely studied in hybrid spin-mechanical system \cite{PhysRevA.107.023722,PhysRevLett.117.015502}, the quantum spin lattice systems \cite{PhysRevLett.94.097203,RN493} and hybrid circuit systems\cite{RN492,PhysRevLett.128.010604}.
In this work, we focus on the entanglement dynamics change induced by the EPs, and the main results are summarized in Fig.~\ref{Fig:1}. 

\section{Energy spectrum and EPs}
\label{3}

\begin{figure}[htbp]
\includegraphics[angle=0,width=1\linewidth]{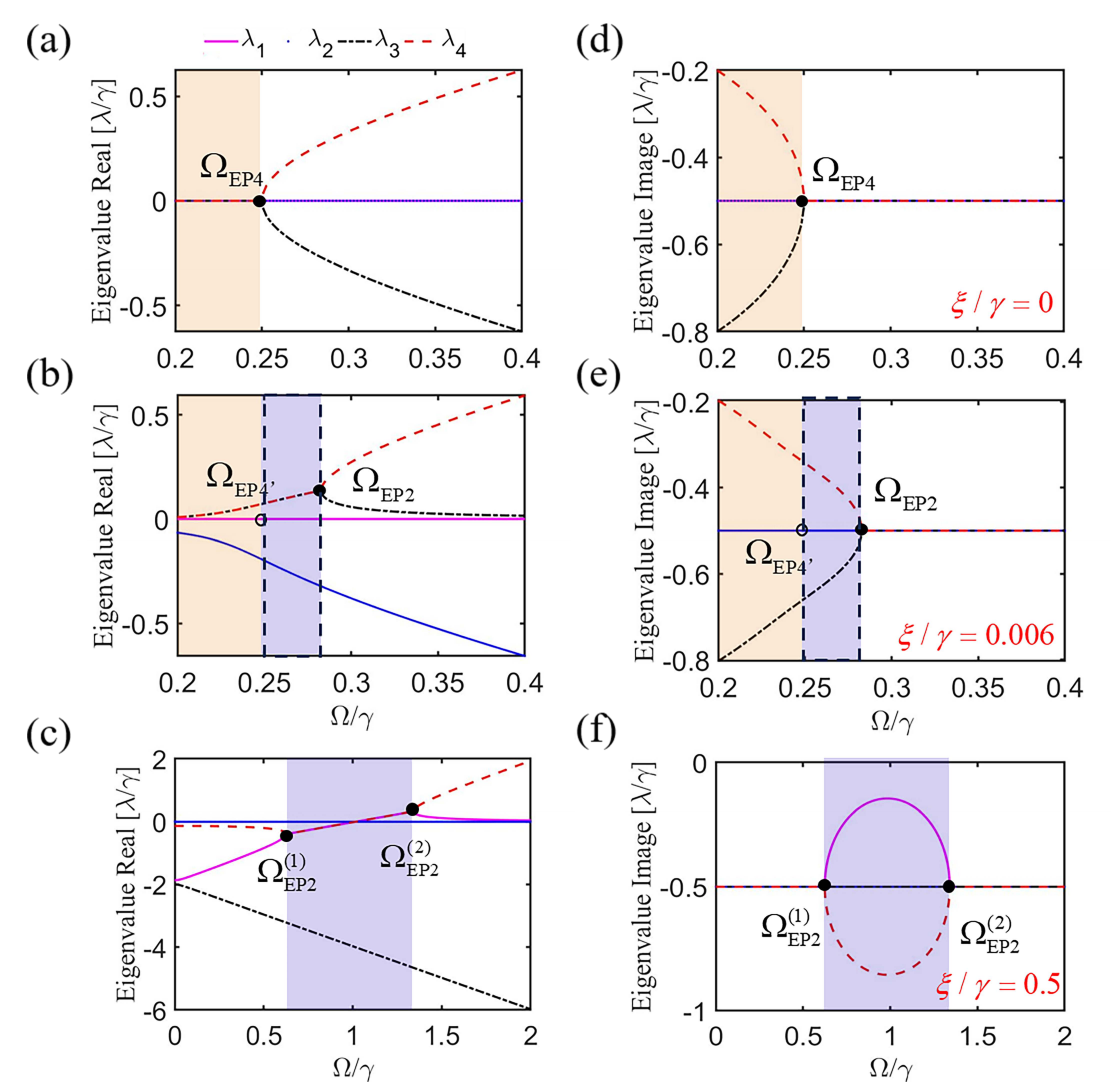}
\caption{
Real (a, b, c) and imaginary (d, e, f) parts of eigenvalues for different Ising type interaction strength. (a) Real and (d) imaginary parts of the eigenvalues for $\xi /\gamma =0$. As examples for weak coupling regime for $\xi /\gamma =0.006$ (b, e) and strong coupling regime for $\xi /\gamma =0.5$ (c, f), respectively.
The two qubits are assumed that have the same drive amplitude $\Omega$ and the same decay rates $\gamma $. The orange and white regions represent the PI and PTS, respectively. The mixed phase is formed by the original EP4 and existing EP2 (i.e., ${{\Omega }_{\text{EP4}'}} <\Omega <{{\Omega }_{\text{EP2}}}$) in the case of weak coupling, while is surrounded by the two EP2s (i.e., $\Omega _{\text{EP2}}^{(1)}<\Omega <\Omega _{\text{EP2}}^{(2)}$) in the strong coupling regime, 
as illustrated in the blue regions.}
\label{Fig:2}
\end{figure}

In the following we mainly focus on the case of Ising type interaction (the calculation of the dipolar interaction case can be found in Sec. \ref{6}).
The system Hamiltonian in Eq.~(\ref{Eq:1}) reduces to a so-called passive $\mathcal{P}\mathcal{T}$-symmetric Hamiltonian \cite{PhysRevLett.103.093902} can be written as $H={{H}_{\mathcal{PT}}}-\left( i\gamma /2 \right)I$ with $I$ being the unitary matrix. The Hamiltonian ${{H}_{\mathcal{PT}}}$ can be written as 
\begin{equation}
    \begin{split}
    {{H}_\mathcal{PT}} = &\left( \begin{matrix}
       -2\xi+\frac{i\gamma }{2} & \Omega  & \Omega  & -2\xi  \\
       \Omega  & -2\xi & -2\xi & \Omega   \\
       \Omega  & -2\xi & -2\xi & \Omega   \\
       -2\xi & \Omega  & \Omega  & -2\xi-\frac{i\gamma }{2}  \\
    \end{matrix} \right).
    \end{split}
    \end{equation}
This Hamiltonian satisfies $\mathcal{P}\mathcal{T}$ symmetry, where the parity operator is given as $\mathcal{P}=\sigma _{1}^{x}\sigma _{2}^{x}$.
The eigenvalues of the  system without coupling, i.e., $\xi =0$ ($g =0$), are ${{\lambda }_{0,1}}=-i\gamma /2$ and ${{\lambda }_{\pm }}=-i\gamma /2\pm i\sqrt{16{{\Omega }^{2}}-{{\gamma }^{2}}}/2$, indicating that the system has a four-order EP (EP4) if $\Omega ={{\Omega }_{\text{EP4}}}=\gamma /4$ [Figs.~\ref{Fig:2}(a) and ~\ref{Fig:2}(d)].
In the region of $\mathcal{P}\mathcal{T}$-symmetric phase (PTS) (i.e., $\Omega >{{\Omega }_{\text{EP4}}}$), the eigenvalues of the system are purely real, while the eigenvalues are purely imaginary (PI) in the region of $\Omega < {{\Omega }_{\text{EP4}}}$, this phase region is defined as PI [see Fig.~\ref{Fig:2}(a, d)].
As described in Fig.~\ref{Fig:2}(b) and ~\ref{Fig:2}(e), an EP2 can be found at ${{\Omega }_{\text{EP2}}}/\gamma =0.283$ when choosing $\xi/\gamma = 0.006$ as an example.  
Although the EP4 disappears in the weak coupling regime (i.e., $\xi \ll \gamma $), it still has influence on the quantum effect. Thus, the original EP4 (i.e., EP4') can be read as the “ghost” of the factual EP4. 
Corresponding to the presence of EP4', the purely imaginary (PI) eigenvalues become a mixture of both real and imaginary parts. The EP2 induces the spectrum transition of the PTS and the mixed phase [see the blue regions in Fig.~\ref{Fig:2}(b, e)]. 
It is well-known that in the region of $\mathcal{P}\mathcal{T}$-symmetry-broken phase (PTB), the eigenvalues are a pair of conjugate complex numbers, that is $\lambda =-{{\lambda }^{*}}$. Therefore, both the mixed phase and PI belong to the PTB.

The order of EP will be lowered from 4 to 2 when adding the qubit-qubit coupling. An additional EP2 will appear in the system when enhancing the coupling strength to the strong coupling regime (i.e., $\xi \sim\gamma $). As demonstrated in Fig.~\ref{Fig:2}(c) and ~\ref{Fig:2}(f), one EP2 appears at the critical value $\Omega _{\text{EP2}}^{(1)}/\gamma =0.626$ and an additional EP2 is at $\Omega _{\text{EP2}}^{(2)}/\gamma =1.34$ when choosing $\xi/\gamma = 0.5$. The region of the mixed phase occurs at $\Omega _{\text{EP2}}^{(1)}<\Omega <\Omega _{\text{EP2}}^{(2)}$ and the regions of PTS take place at $\Omega <\Omega _{\text{EP2}}^{(1)}$ and $\Omega >\Omega _{\text{EP2}}^{(2)}$, which indicates that the mixed phase is surrounded by the two EP2s in the strong coupling regime.

\section{Different entanglement dynamics}
\label{4}

The state evolution of the coupled NH qubits can be obtained by solving the equation of motion
\begin{equation}
\frac{d\rho }{dt}=-i\left( {{H}_{j=1,2}}\rho -\rho H_{j=1,2}^{\dagger } \right),
\label{Eq:2}
\end{equation}
where $\rho $ denotes the density matrix of the system. 
Concurrence is a useful tool for quantifying the entanglement of bipartite quantum systems, which can be written as \cite{PhysRevLett.80.2245}
\begin{equation}
C=\max \left( {{\tau }_{1}}-{{\tau }_{2}}-{{\tau }_{3}}-{{\tau }_{4}},0 \right).
\label{Eq:3}
\end{equation}
Here ${{\tau }_{1}}$, ${{\tau }_{2}}$, ${{\tau }_{3}}$, ${{\tau }_{4}}$ are the eigenvalues of the Hermitian matrix $R=\sqrt{\sqrt{\rho }\widetilde{\rho }\sqrt{\rho }}$, with  $\widetilde{\rho }=\left( {{\sigma }_{y}}\otimes {{\sigma }_{y}} \right){{\rho }^{*}}\left( {{\sigma }_{y}}\otimes {{\sigma }_{y}} \right)$ in decreasing order, 
${{\rho }^{*}}$ is the complex conjugation of the density matrix $\rho $ and ${{\sigma }_{y}}=-i\left| b \right\rangle \left\langle  a \right|+i\left| a \right\rangle \left\langle  b \right|$.

\begin{figure}[htbp]
\includegraphics[angle=0,width=1\linewidth]{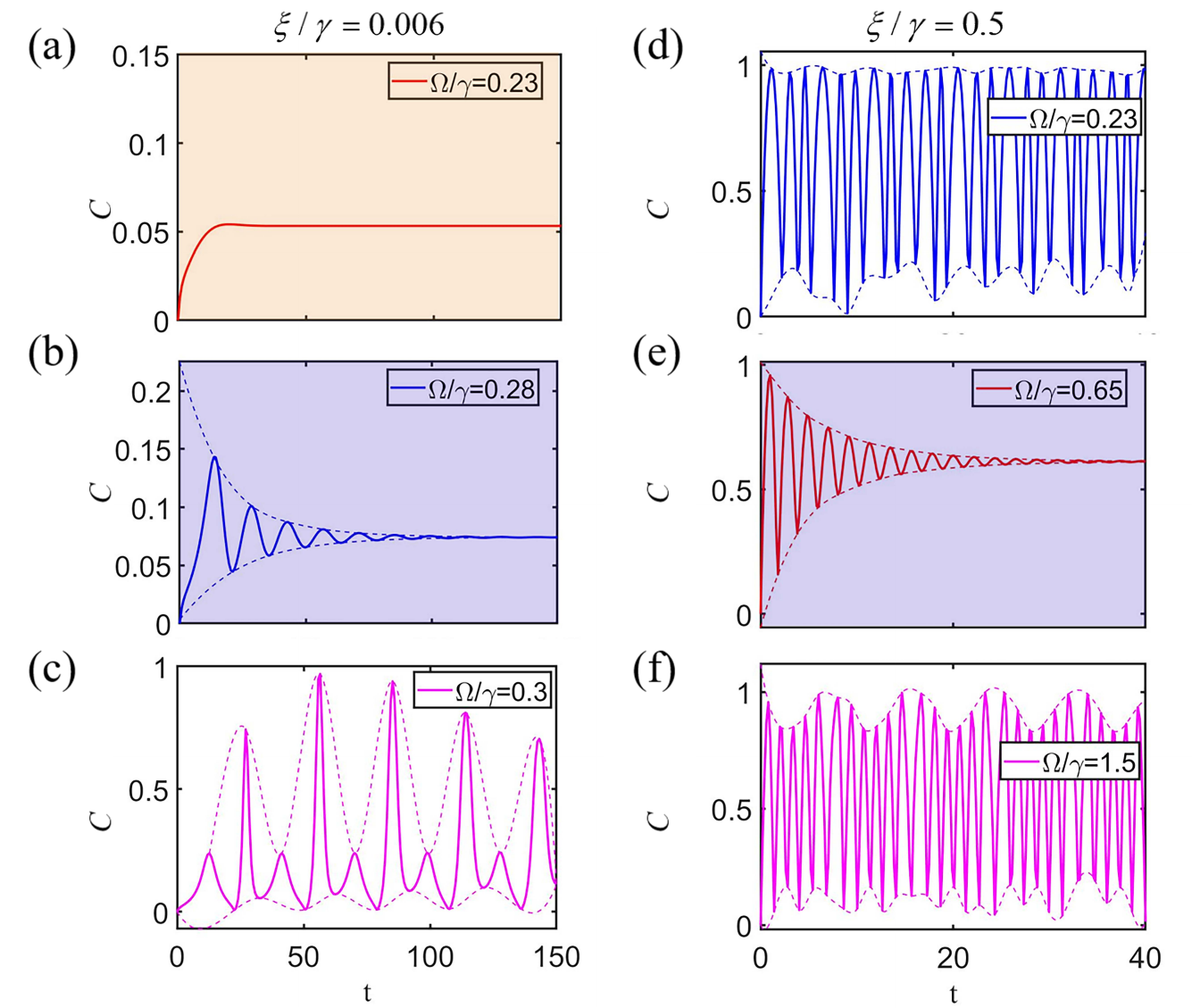}
\caption{
  As an example for weak coupling strength $\xi/\gamma = 0.006$, the concurrence evolution in the region of PI (a), the mixed phase (b) and PTS (c). The original EP4 and EP2 induce the entanglement dynamics change on the two sides of these points.
  In the strong coupling regime for $\xi/\gamma = 0.5$, the EP2 induces the spectrum transition of the PTS and mixed phase, leading to the type II (d, f) and III (e) entanglement dynamics change. 
  All plots are made for $\left| \psi (0) \right\rangle =\left| aa \right\rangle $. The dashed curves denote the envelope fitting of the concurrence evolutions. 
}
\label{Fig:3}
\end{figure}

For a given initial state $\left| \psi (0) \right\rangle $, the normalized quantum state can be expressed as $\left| \widetilde{\psi }(t) \right\rangle =\alpha (t)\left| aa \right\rangle +\beta (t)\left| ab \right\rangle +\zeta (t)\left| ba \right\rangle +\delta (t)\left| bb \right\rangle $. Hence, the concurrence is given by
\begin{equation}
C=2\left| \alpha \delta -\beta \zeta  \right|. 
\label{Eq:1111}
\end{equation}
Hence, the concurrence can be used to calculate the degree of entanglement between the qubits (more details are given in Appendix~\ref{Sec:A5}).

By solving Eq.~(\ref{Eq:2}) and Eq.~(\ref{Eq:3}), the numerical results of the concurrence evolution are displayed in Fig.~\ref{Fig:3}.
In the case of weak coupling (taking $\xi/\gamma = 0.006$ as an example), the entanglement dynamics show exponential-like behavior in the region of $\Omega <{{\Omega }_{\text{EP4}'}}$ [Fig.~\ref{Fig:3}(a)]. We define this type of entanglement dynamics as type I, which can only be observed in the weak coupling regime. 
In the region of PTS, i.e., $\Omega >{{\Omega }_{\text{EP2}}}$, the entanglement dynamics exhibits continuous oscillation in the whole time-evolution process [Fig.~\ref{Fig:3}(c)], which is defined as type II. By fitting envelope functions to the type II, the most obvious characteristic of this type entanglement is that both the upper and lower branches exhibit significant oscillatory behavior.
In the mixed phase, the feature of the entanglement dynamics exhibits a combination of both the exponential-like and oscillatory behavior [Fig.~\ref{Fig:3}(b)]. Clearly, this type of entanglement dynamics is defined as type III. One can distinguish the type III entanglement dynamics in the strong coupling and weak coupling regimes by comparing the decay rate of the upper and lower envelopes (see Fig.~\ref{Fig:S12}).

Meanwhile, the concurrence evolution is ranging from 0 to 1, and the maximum entanglement ($C=1$) in PTS is much larger than that in PI and the mixed phase. 
By using the theory of time-independent perturbation theory for non-Hermitian systems \cite{PhysRevLett.131.100202}, we can obtain the analytical results of the concurrence in the weak coupling regime, which shows a good agreement with the numerical results in the $\mathcal{P}\mathcal{T}$-symmetric phase (see Appendix~\ref{Sec:A5} for more details). The different features of the entanglement dynamics also provides an efficient approach to distinguish the spectrum phase (PTS, PTB) of the system.

\begin{figure}[htbp]
\centering
\includegraphics[angle=0,width=1\linewidth]{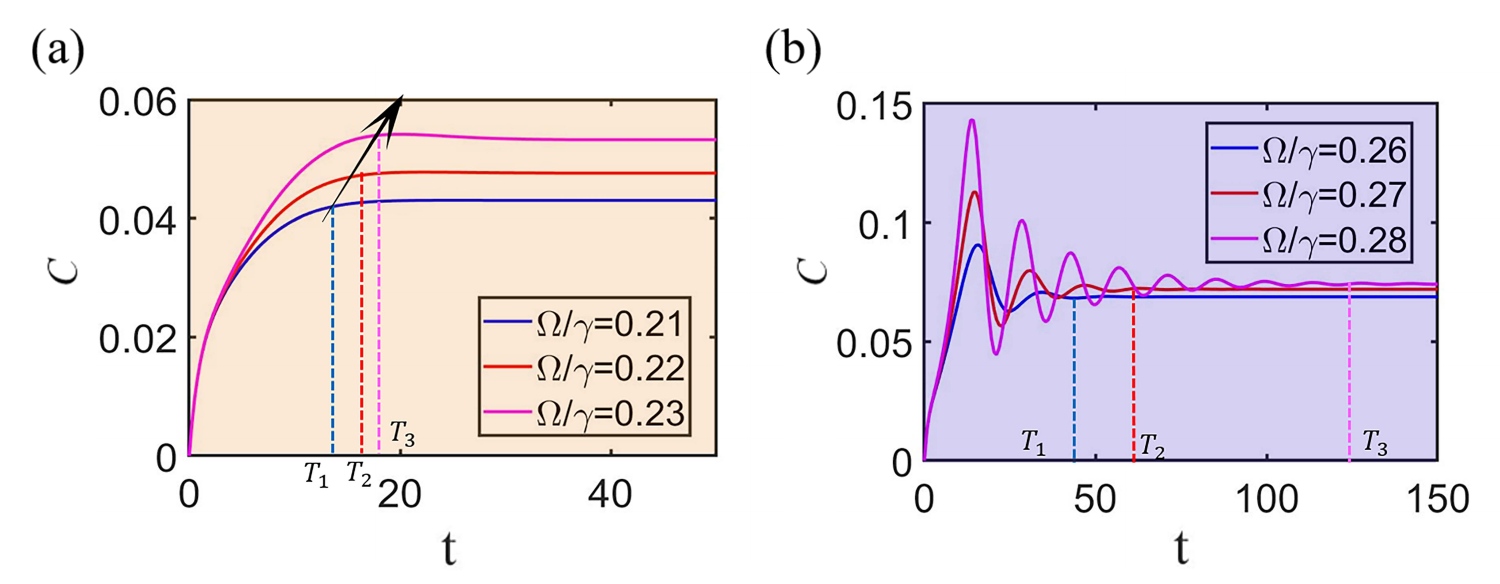}
\caption{
 The concurrence evolution in the PI (a) and the mixed phase (b). In PI, the observation of steady-state entanglement generation at a longer timescale by proximity to the original EP4, as the black arrow shown. However, in the mixed phase, the steady-state entanglement can be generated at a speed much faster when approach to the original EP4. The other parameter is chosen as $\xi /\gamma =0.006$ and the initial state is $\left| \psi (0) \right\rangle =\left| aa \right\rangle $.
}
\label{Fig:S11}
\end{figure}

Furthermore, we display that on approaching the EP4’ in PI (i.e., $\Omega <{{\Omega }_{\text{EP}4'}}$), the speed of reaching steady-state entanglement decreases, meaning that the required time becomes longer, i.e., ${{T}_{1}}<{{T}_{2}}<{{T}_{3}}$ (see Fig.~\ref{Fig:S11}(a)). On the contrary, in the mixed phase (i.e., ${{\Omega }_{\text{EP}4'}}<\Omega <{{\Omega }_{\text{EP2}}}$), when forcing proximity to the EP4’, steady-state entanglement can be achieved more quickly (see Fig.~\ref{Fig:S11}(b)). Based on this observation, although both the PI and the mixed phase fall within the $\mathcal{P}\mathcal{T}$-symmetry-broken phase, the steady-state entanglement generation near higher-order EPs behaves in completely opposite ways in these two regions.

\FloatBarrier

This phenomenon can be understood as follows. On the one hand, if the applied driving amplitude smaller than the original EP4, i.e., $\Omega <{{\Omega }_{\text{EP}4'}}$, even a smaller driving amplitude can lead to realize faster steady-state entanglement generation. On the other hand, once the driving amplitude larger than the original EP4, i.e., $\Omega >{{\Omega }_{\text{EP}4'}}$, where the EP4' can be treated as a perturbation in the parameter space. The greater the driving amplitude, the larger the oscillations in the concurrence evolution, resulting in a longer timescale to achieve steady-state entanglement.

With the increase of the coupling strength to the strong coupling regime, i.e., $\xi/\gamma = 0.5$ as an example, the original EP4 is replaced by another EP2, which means there are two different EP2s in the system [see Figs.~\ref{Fig:2}(c) and ~\ref{Fig:2}(f)]. The entanglement between two coupled NH qubits has been shown in Figs.~\ref{Fig:3}(d) ~\ref{Fig:3}(f). We can observe that the entanglement dynamics of type II takes place in the region of $\Omega <\Omega _{\text{EP2}}^{(1)}$ and $\Omega >\Omega _{\text{EP2}}^{(2)}$, i.e., the region of PTS, as shown in Figs.~\ref{Fig:3}(d) and ~\ref{Fig:3}(f), respectively. 
In the region of mixed phase, i.e., $\Omega _{\text{EP2}}^{(1)}<\Omega <\Omega _{\text{EP2}}^{(2)}$, the type III entanglement dynamics can be observed, as shown in Fig.~\ref{Fig:3}(e).
Therefore, only two types of entanglement dynamics can be found in the strong coupling regime.

\section{The discussion of the entanglement dynamics change}
\label{5}

The normalized quantum state evolved under the system Hamiltonian Eq.~(\ref{Eq:1}) in the main text can be written as 
\begin{equation}
\left| \widetilde{\psi }(t) \right\rangle =\sum\limits_{n}{{{c}_{n}}{{e}^{-i{{\lambda }_{n}}t}}}\left| {{\lambda }_{n}} \right\rangle and {{\lambda }_{n}}=\operatorname{Re}({{\lambda }_{n}})+i\operatorname{Im}({{\lambda }_{n}}), 
\label{Eq:111}
\end{equation}
where ${{\lambda }_{n}}$ and $\left| {{\lambda }_{n}} \right\rangle $ are eigenvalues and eigenstates of the non-Hermitian Hamiltonian Eq.~(\ref{Eq:1}), respectively. Given an initial state $\left| \psi (0) \right\rangle $, the time evolution of the state under the system Hamiltonian $U={{e}^{-iHt}}$ can be read as
\begin{equation}
\left| \widetilde{\psi }(t) \right\rangle =U\left| \psi (0) \right\rangle =\sum\limits_{n=1}^{4}{\left\langle  {{\lambda }_{n}} | \psi (0) \right\rangle \cdot {{e}^{-it{{\lambda }_{n}}}}\left| {{\lambda }_{n}} \right\rangle }, 
\label{Eq:222}
\end{equation}

Correspondingly, we can obtain the coefficient of the state as 
$\alpha (t)={{c}_{1}}\cdot {{e}^{-it{{\lambda }_{1}}}},\beta (t)={{c}_{2}}\cdot {{e}^{-it{{\lambda }_{2}}}},\varsigma (t)={{c}_{3}}\cdot {{e}^{-it{{\lambda }_{3}}}},\delta (t)={{c}_{4}}\cdot {{e}^{-it{{\lambda }_{4}}}}$,
with the coefficients determined by the initial state as
${{c}_{1}}=\left\langle  {{\lambda }_{1}} | \psi (0) \right\rangle$, ${{c}_{2}}=\left\langle  {{\lambda }_{2}} | \psi (0) \right\rangle$, ${{c}_{3}}=\left\langle  {{\lambda }_{3}} | \psi (0) \right\rangle$ and ${{c}_{4}}=\left\langle  {{\lambda }_{4}} | \psi (0) \right\rangle$. In this case, the concurrence evolution can be given by
\begin{align}
  C(t) &= 2\left| c_{1} c_{4} e^{-it(\lambda_{1} + \lambda_{4})} - c_{2} c_{3} e^{-it(\lambda_{2} + \lambda_{3})} \right| \notag \\ 
       &= 2\left| c_{1} c_{4} e^{-it \operatorname{Re}(\lambda_{1} + \lambda_{4})} - c_{2} c_{3} e^{-it \operatorname{Re}(\lambda_{2} + \lambda_{3})} \right| \notag \\ 
       &\quad + 2\left| c_{1} c_{4} e^{t \operatorname{Im}(\lambda_{1} + \lambda_{4})} - c_{2} c_{3} e^{t \operatorname{Im}(\lambda_{2} + \lambda_{3})} \right|.
\label{Eq:333}
\end{align}

It can be seen that the imaginary part $\operatorname{Im}({{\lambda }_{n}})$ will cause the concurrence evolution exhibiting exponential-like behavior, while the real part $\operatorname{Re}({{\lambda }_{n}})$ will cause the concurrence evolution showing oscillatory behavior. 

According to Eq.~(\ref{Eq:333}), we can find that the eigenvalues of the system undergoes a transition at EPs from purely real to purely imaginary or complex, the evolution of the concurrence also changes. Specifically, we compare the different characteristic of eigenvalues in three regions (i.e., the PI, the mixed phase and the PTS). The original EP4 induces the spectrum transition of the PI and mixed phase, leading to the type I (exponentially-like behavior, see Fig.~\ref{Fig:3}(a)) and III (combination of both the exponential-like decay and oscillatory behavior, see Fig.~\ref{Fig:3}(b, e)) entanglement dynamics change in the weak coupling regime. Similarly, the EP2 induces the spectrum transition of the PTS and mixed phase, leading to the type II (oscillatory behavior, see Fig.~\ref{Fig:3}(c, d, f)) and III entanglement dynamics change. This effect can be observed in both the weak and strong coupling regimes.

\begin{figure}[htbp]
\includegraphics[angle=0,width=1\linewidth]{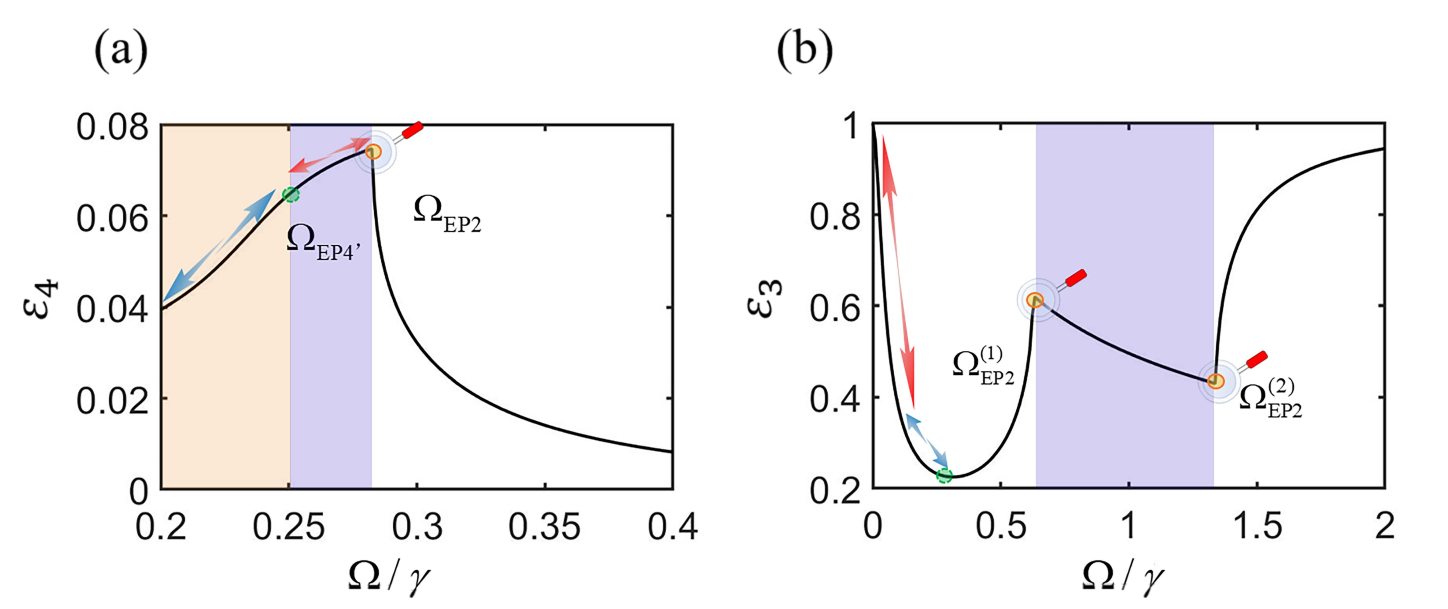}
\caption{
 The concurrence value ${{\varepsilon }_{4}}$ in the weak coupling regime (a) and ${{\varepsilon }_{3}}$ in the strong coupling regime (b) as a function of the driving amplitude $\Omega $. The yellow circles represent the EP2 in the system while the green circles denote the original EP4. }
\label{Fig:S9}
\end{figure}

As the eigenstates are entangled states, the concurrence about eigenstates can be treated as another manifestation of the entanglement dynamics change [see Fig. ~\ref{Fig:S9}]. By giving concurrence associated with the fourth eigenstate ${{\varepsilon }_{4}}$ in the weak coupling regime [Fig. ~\ref{Fig:S9}(a)] and the third eigenstate ${{\varepsilon }_{3}}$ [Fig. ~\ref{Fig:S9}(b)] in the strong coupling regime, we demonstrate that there is a notable inflection point appears in the concurrence evolution at the EP2. In particular, the slope of the concurrence evolution changes abruptly at the inflection point (i.e., EP2), suggesting that the occurrence of an entanglement dynamics change at the EP2. In addition, we note that there is a notable change in the trend of the concurrence evolution near the EP4’ in both weak and strong coupling regimes, as the red and blue arrows shown. As a result, the entanglement dynamics change between type I and II can be observed at the EP4’.

\section{Entanglement changing without the driving field}
\label{6}

\begin{figure}[htbp]
\includegraphics[angle=0,width=1\linewidth]{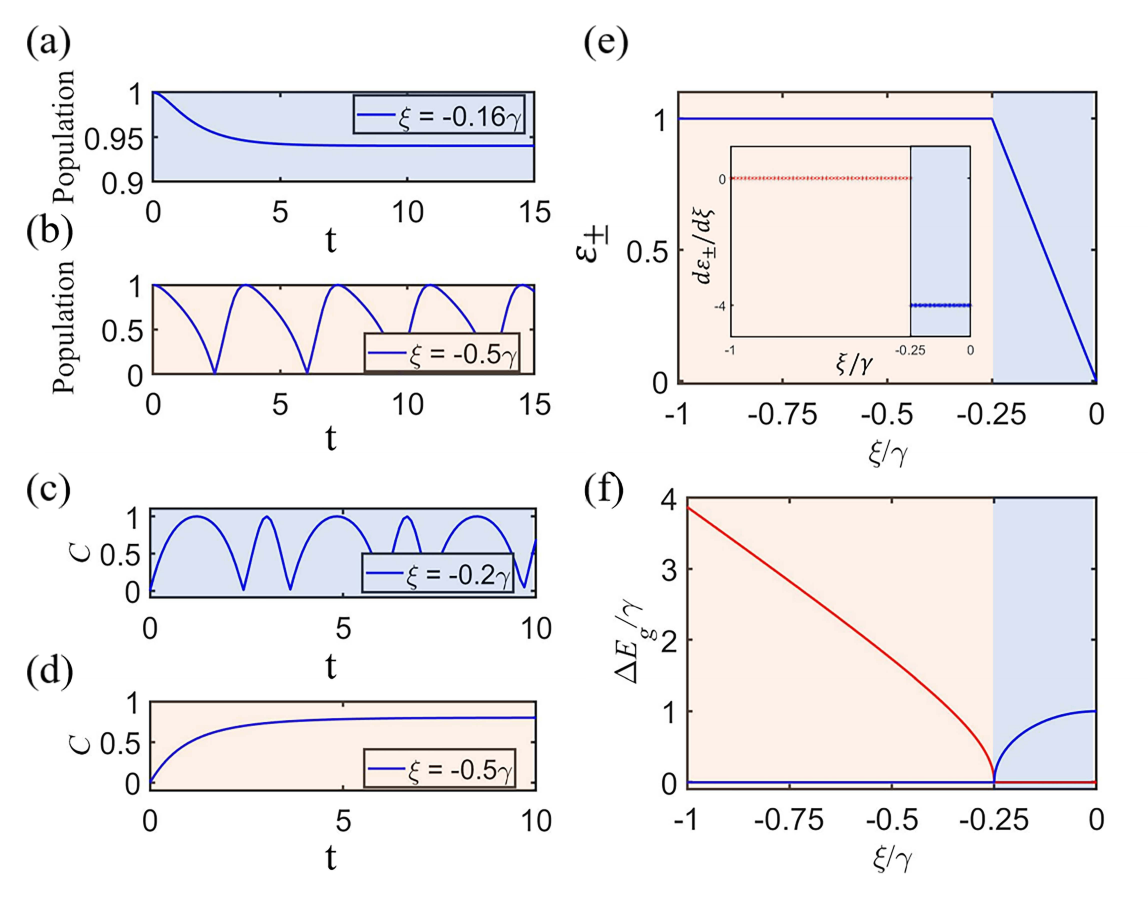}
\caption{
Observation of exceptional phase transitions in the absence of $\Omega $. (a) and (b) Measured evolution of the population of the basis state $\left| {{b}_{1}},{{b}_{2}} \right\rangle $ in PTB (for $\xi /\gamma =-0.16$) and PTS (for $\xi /\gamma =-0.5$), respectively. Concurrence evolution are given for $\xi =-0.2\gamma $ (c) and $\xi =-0.5\gamma $ (d). (e) Concurrence values ${{\varepsilon }_{\pm }}$ for the eigenstates $\left| {{\Phi }_{+}} \right\rangle $ and $\left| {{\Phi }_{-}} \right\rangle $ versus the Ising type interaction strength $\xi $. The derivatives $d{{\varepsilon }_{\pm }}/d\xi $ around the EP2, obtained by $\left[ {{\varepsilon }_{\pm }}(\xi +\delta \xi )-{{\varepsilon }_{\pm }}(\xi ) \right]/d\xi $, as shown in the insets. (f) Spectral gap $\Delta {{E}_{g}}$. The eigenspectrum is possessed the entangled eigenstates of the two qubits system. This gap corresponds to the vacuum Rabi splitting.
}
\label{Fig:4}
\end{figure}

In the absence of the driving field applied to the qubits, i.e., ${{\Omega }_{i=1,2}}=0$. Exceptional entanglement changing can also occur when considering the Ising type interaction.
The Hamiltonian with Ising type interaction evolves within the subspace $\left\{ \left| {{a}_{1}},{{a}_{2}} \right\rangle ,\left| {{a}_{1}},{{b}_{2}} \right\rangle ,\left| {{a}_{2}},{{b}_{1}} \right\rangle ,\left| {{b}_{1}},{{b}_{2}} \right\rangle  \right\}$. 
With a correction of the quantum state distortion caused by the decoherence, the evolution of the joint probability $\left| {{b}_{1}},{{b}_{2}} \right\rangle $, denoted as ${{P}_{{{b}_{1}}{{b}_{2}}}}$, indicating the quantum Rabi oscillator signal. Figure~\ref{Fig:4}(a) and ~\ref{Fig:4}(b) show the obtained ${{P}_{{{b}_{1}}{{b}_{2}}}}$ as functions of time in the regions of different phases. Specifically, in PTB (e.g., $\xi/\gamma = -0.16$), the population of state $\left| {{b}_{1}},{{b}_{2}} \right\rangle $ monotonically decays during the system evolving to a steady state [Fig.~\ref{Fig:4}(a)].
After crossing the EP, i.e., in the region of PTS (e.g., $\xi/\gamma = -0.5$), the state population ${{P}_{{{b}_{1}}{{b}_{2}}}}$ presents a distinct oscillatory behavior in time evolution as shown in Fig.~\ref{Fig:4}(b).

To show the relationship between entanglement behavior and the EP, we give the concurrence evolution for $\xi =-0.2\gamma $ and $-0.5\gamma $ in Fig.~\ref{Fig:4}(c) and ~\ref{Fig:4}(d), respectively. This result demonstrates that the entanglement exhibits different evolution behaviors in PTB and PTS.
Figure~\ref{Fig:4}(e) indicates the concurrence values ${{\varepsilon }_{\pm }}$ associated with the two eigenstates $\left| {{\Phi }_{\pm }} \right\rangle $ of the system as functions of $\xi $. As theoretical predicted, below the EP (i.e., $\left| \xi /\gamma  \right|>{\gamma }/{4}\;$), the concurrence value of each eigenstate is saturated and approximately converges to the same maximally entangled state (reaching the maximum value 1) and independent of $\xi $ until at the EP $\xi =-\gamma /4$, where the energy gap disappeared. The expression of the concurrence has been given in Appendix~\ref{Sec:A0}.
The discontinuity of these derivatives indicates the occurrence of an entanglement changing at the EP, which are shown in the insets.
After crossing the EP (i.e., $\left| \xi /\gamma  \right|<{\gamma }/{4}\;$), the concurrence values ${{\varepsilon }_{\pm }}$ exhibit a linear scaling and depends on $\xi $. ${{\varepsilon }_{\pm }}$ are decreasing to 0 with the reduction of $\left| \xi  \right|$.
In Fig.~\ref{Fig:4}(f), the energy gap described as ${{E}_{g}}=\sqrt{4{{J}^{2}}-{{\gamma }^{2}}}$ with $J=-2\xi$, undergoes a transition from the real part to the imaginary part at the EP, which is accompanied by an entanglement transition of the eigenstates. In addition, according to the results in Figs.~\ref{Fig:4}(a),~\ref{Fig:4}(b) and ~\ref{Fig:4}(e), this type of the energy gap transition is also accompanied by the effect of the vacuum Rabi splitting of the entangled states.

\section{The calculation for the system with dipolar interaction}
\label{7}

In addition to the case of the Ising type interaction discussed above, we also considered the dipolar interaction between the qubits, which can be given by 
\begin{equation}
{{H}_{\operatorname{int}}}=g(\sigma _{1}^{+}\sigma _{2}^{-}+\sigma _{1}^{-}\sigma _{2}^{+}).
\label{Eq:S9}
\end{equation}
We defined $g$ as the effective dipolar interaction between the two qubits. In this case, the matrix of the total Hamiltonian can be written as
\begin{equation}
H=\begin{pmatrix}
0 & \Omega & \Omega & 0 \\
\Omega & -i\frac{\gamma}{2} & g & \Omega \\
\Omega & g & -i\frac{\gamma}{2} & \Omega \\
0 & \Omega & \Omega & g-i\gamma \\
\end{pmatrix}.
\label{Eq:S10}
\end{equation}

\begin{figure}[htbp]
\includegraphics[angle=0,width=1\linewidth]{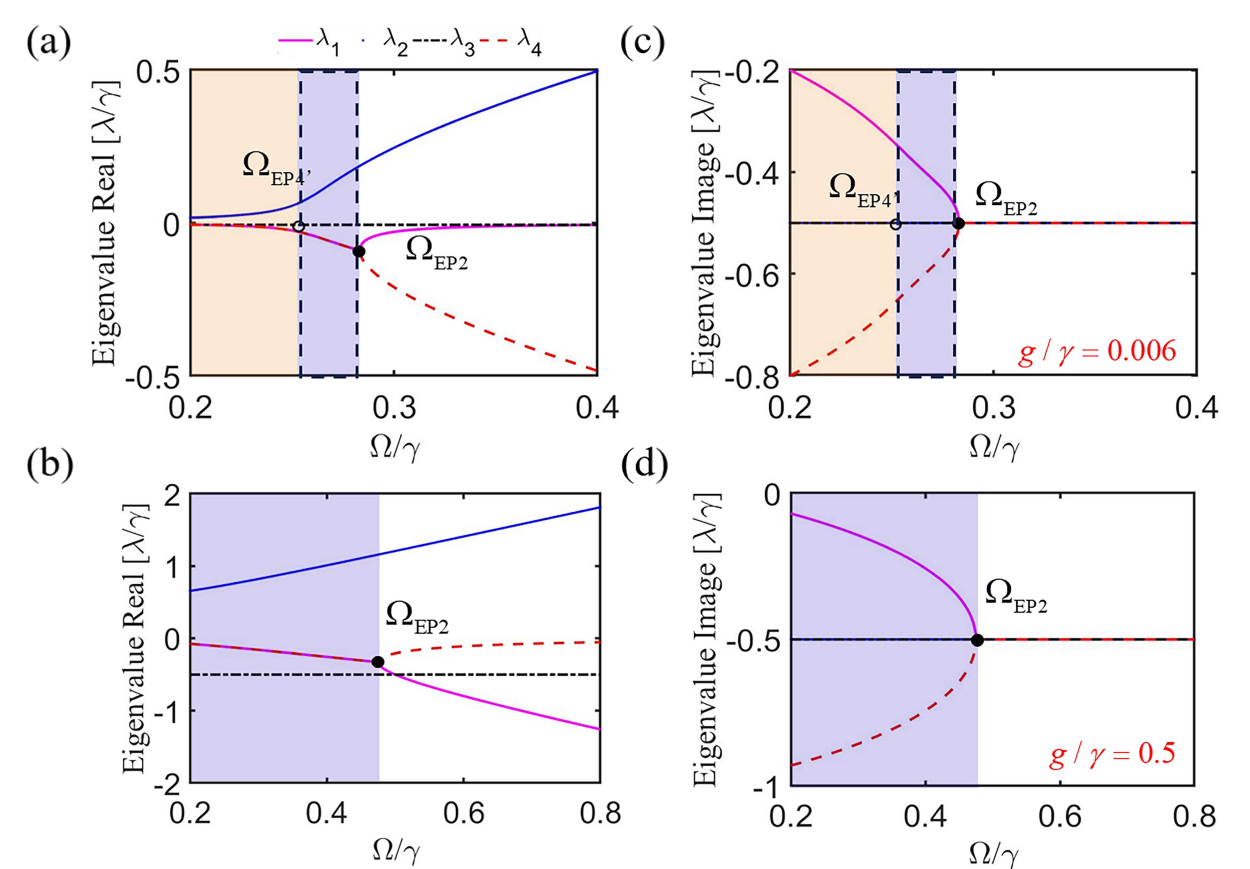}
\caption{
Real (a, b) and imaginary (c, d) parts of eigenvalues for different dipolar coupling regimes as $g /\gamma =0.006$ (a, c) in the weak coupling regime (i.e., $g\ll \gamma $) and $g /\gamma =0.5$ (b, d) in the strong coupling regime (i.e., $g \sim\gamma $), respectively. The two qubits are assumed to have the same drive amplitude $\Omega$ and the same decay rates$\gamma$. The orange, blue and white regions represent the PI, the mixed phase and PTS, respectively. The value of original EP4 is at the critical point ${{\Omega }_{\text{EP4}'}}={{\Omega }_{\text{EP4}}}=\gamma /4$.
}
\label{Fig:S4}
\end{figure}

We now investigate the relationship with the EP properties and entanglement behavior of the system, where the two qubits are coupled with dipolar coupling. By numerically solving the matrix of the Hamiltonian ~(\ref{Eq:S10}), we can obtain the eigenvalues of the system, as shown in Figure~\ref{Fig:S4}. In the weak coupling regime (i.e., $g \ll \gamma $) for $g/\gamma = 0.006$, there exists an EP2 at $\Omega/\gamma = 0.283$ [Figs.~\ref{Fig:S4}(a) and ~\ref{Fig:S4}(c)]. The region of PI is in $\Omega <{{\Omega }_{\text{EP4}'}}$ while the mixed phase is in the region of ${{\Omega }_{\text{EP4}'}}<\Omega <{{\Omega }_{\text{EP2}}}$. Similarly, both the PI and mixed phase belong to the PTB. 
Only a single EP2 appears at $\Omega/\gamma = 0.47$ in the strong coupling regime (i.e., $g \sim\gamma $) for $g/\gamma = 0.5$ [Fig.~\ref{Fig:S4}(b) and ~\ref{Fig:S4}(d)]. 

\begin{figure}[htbp]
\includegraphics[angle=0,width=1\linewidth]{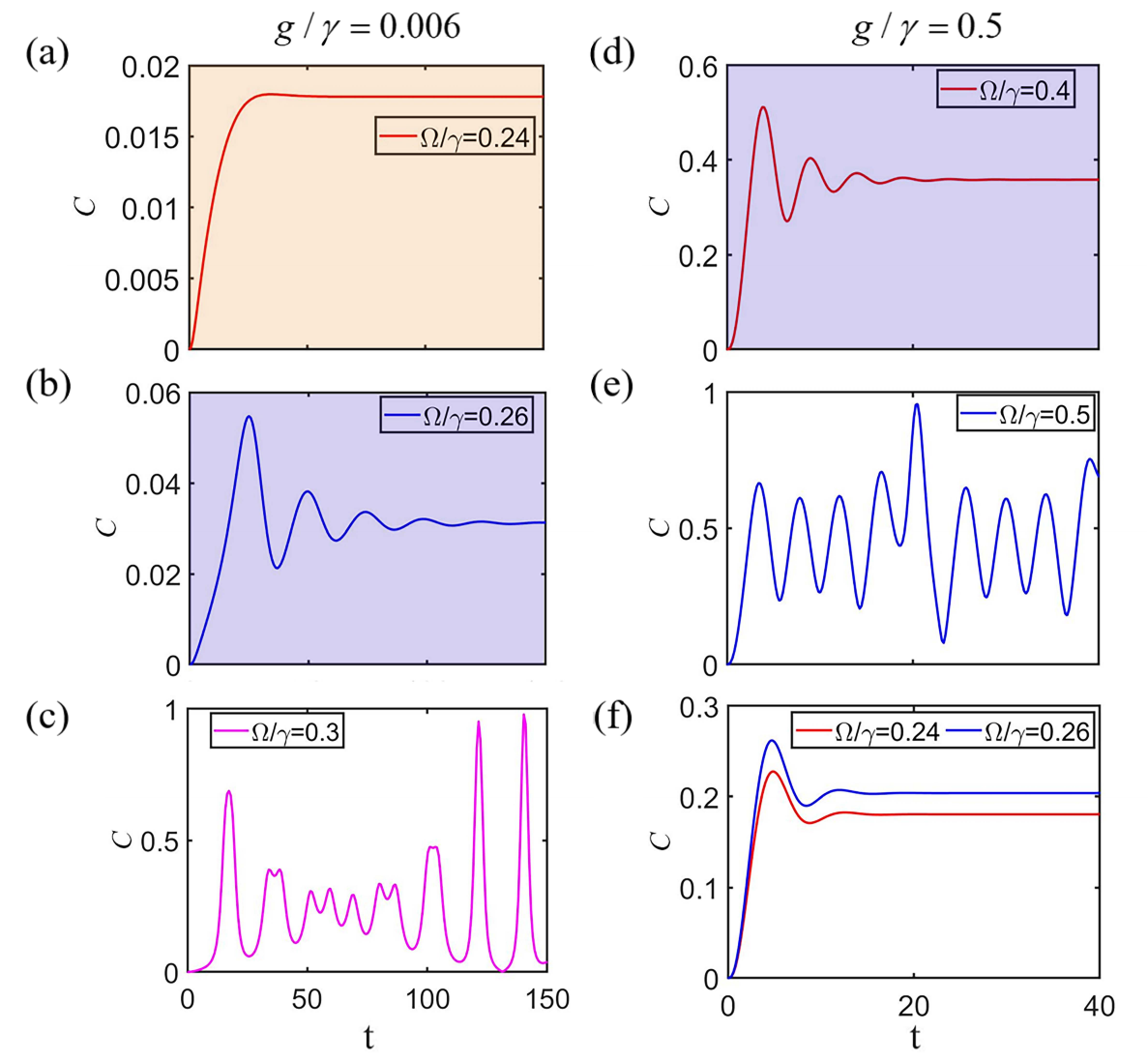}
\caption{
  For the case of dipolar interaction, the concurrence evolution in PI (a), the mixed phase (b) and PTS (c) for weak coupling strength $\xi/\gamma = 0.006$. The EP4' and EP2 induce the different entanglement dynamics changes on the two sides of these points.
  In the strong coupling regime for $\xi/\gamma = 0.5$, the EP2 induces the spectrum transition of the PTS and mixed phase, leading to the type II (e) and III (d) entanglement dynamics change. The entanglement dynamics for the driving amplitude $\Omega/\gamma = 0.24, 0.26$ (f).
  All the initial states are $\left| \psi (0) \right\rangle =\left| aa \right\rangle $.
}
\label{Fig:S6}
\end{figure}

In the weak coupling regime, we can still observe the different entanglement dynamics change at the EP4' and EP2, similarly with the model of Ising type interaction system, as shown in Fig.~\ref{Fig:S6}(a)-Fig.~\ref{Fig:S6}(c). Correspondingly, the characteristic of the entanglement dynamics also depends on the feature of the eigenvalues.  

In the strong coupling regime, the EP2 induces the spectrum transition of the PTS and mixed phase, leading to the type II [Fig.~\ref{Fig:S6}(e)] and III [Fig.~\ref{Fig:S6}(d)] entanglement dynamics change.
However, the influence of original EP4 on entanglement behavior is disappears, indicating that only type III entanglement dynamics can be observed, as shown in Fig.~\ref{Fig:S6}(f). Similarly, as in the case of Ising type interaction, type I entanglement dynamics only occur in the weak coupling regime

\section{Conclusions}
\label{8}

In conclusion, we have found that entanglement dynamics change can be induced by the EPs in a system of two coupled NH qubits. Divided by EP2 and the original EP4, three distinct types of entanglement dynamics can be observed. By analyzing the real and imaginary parts of the eigenvalues of the system, we demonstrate that the characteristics of the eigenvalues determine the types of the entanglement dynamics.
The original EP4 induces the spectrum transition of the PI and mixed phase, leading to the type I and III entanglement dynamics change in the weak coupling regime. Similarly, the EP2 induces the spectrum transition of the PTS and mixed phase, leading to the type II and III entanglement dynamics change in the both weak and strong coupling regimes.
Furthermore, as the eigenstates are entangled states, we further confirm that the concurrence associated with the eigenstates can be treated as another manifestation of the entanglement dynamics change.
Moreover, we have also proved that an EP-enabled entanglement changes in the system without the driving amplitude in the case of Ising-type interaction. Our work provides a protocol for exploring the connection between the entanglement and higher-order EPs, offering opportunities for designing quantum devices by engineering EPs.

\begin{appendix}
    \renewcommand{\thefigure}{\thesection\arabic{figure}} % Change figure numbering in appendix
    
    \section{The concurrence for the two qubits without the driving amplitude}
    \label{Sec:A0}
    For the eigenstates $\left| {{\Phi }_{\pm }} \right\rangle $ of the Hamiltonian without the driving term, the system density operator in the basis states $\left| {{b}_{1}},{{b}_{2}} \right\rangle$ ,$\left| {{b}_{1}},{{a}_{2}} \right\rangle ,\left| {{a}_{1}},{{b}_{2}} \right\rangle ,\left| {{a}_{1}},{{a}_{2}} \right\rangle $ can be expressed as
    
    \begin{equation}
    {{\rho }_{\pm }}={{\left| {{N}_{\pm }} \right|}^{2}}\left( \begin{matrix}
       {{J}^{2}} & 0 & 0 & J{{\Gamma }_{\pm }}  \\
       0 & 0 & 0 & 0  \\
       0 & 0 & 0 & 0  \\
       J\Gamma _{\pm }^{*} & 0 & 0 & {{\left| {{\Gamma }_{\pm }} \right|}^{2}}  \\
    \end{matrix} \right).
    \label{Eq:4}
    \end{equation}
    
    The dynamics of the system is restricted within the subspace $\left\{ \left| {{b}_{1}},{{b}_{2}} \right\rangle ,\left| {{a}_{1}},{{a}_{2}} \right\rangle  \right\}$.
    In such a subspace, the eigenstates of the NH Hamiltonian are obtained by
    \begin{equation}
    \left| {{\Phi }_{\pm }} \right\rangle ={{N}_{\pm }}\left( J\left| {{a}_{1}},{{a}_{2}} \right\rangle +{{\Gamma }_{\pm }}\left| {{b}_{1}},{{b}_{2}} \right\rangle  \right),
    \end{equation}
    where $J=-2\xi$, $N_{\pm}=\left( J^2+|\Gamma_{\pm}|^2 \right)^{-1/2}$ and $\Gamma_{\pm}=i\gamma/2 \pm \sqrt{4J^2-\gamma^2}/2$.
    The energy of these two eigenstates is ${{E}_{g}}=\sqrt{4{{J}^{2}}-{{\gamma }^{2}}}$.
    
    For the case of the two qubits share the same decaying rates, the energy gap undergoes a real-to-imaginary transition at the EP $\xi =-\gamma /4$, which is accompanied by an entanglement transition of the eigenstates.
    The resulting concurrence for the two eigenstates $\left| {{\Phi }_{\pm }} \right\rangle $ are
    \begin{equation}
    {{\varepsilon }_{\pm }}=\frac{2J\left| {{\Gamma }_{\pm }} \right|}{{{J}^{2}}+{{\left| {{\Gamma }_{\pm }} \right|}^{2}}}.
    \label{Eq:5}
    \end{equation}
    With the increase of the Ising interaction strength $\left| \xi  \right|$ until reaching the EP, the energy gap vanishes and both eigenstates approximately converge to the same maximally entangled state as
    \begin{equation}
    \left| \Phi_{\pm} \right\rangle = \frac{1}{\sqrt{2}} \left( \left| a_1, a_2 \right\rangle + i \left| b_1, b_2 \right\rangle \right).
    \label{Eq:6}
    \end{equation}
    
    In the main text, we can observe the different entanglement dynamics around the EP. Physically speaking, these two different behaviors induced by the EP can be understood as a consequence of the competition between the Ising-type interaction and incoherent dissipation.

    \section{Example for the experimental implementation of Ising type interaction}
    \label{Sec:A1}
    \setcounter{figure}{0} % Reset figure counter for each appendix section
    In our system, we consider the hybrid spin-mechanical system, as shown in Fig.~\ref{Fig:S8}(a), where the separated two NV centers are magnetically coupled to the same mechanical motion of a cantilever with dimensions $\left( l,w,t \right)$ via the sharp magnet tip attached to it \cite{PhysRevA.107.023722,PhysRevLett.117.015502,PhysRevLett.125.153602}. By applying the cantilever to a periodic drive that modulates its spring constant \cite{PhysRevLett.67.699}, it is possible to amplify the zero-point fluctuations of the mechanical motion. This phenomenon can be experimentally achieved by situating an electrode close to the lower surface of the cantilever and applying an adjustable, time-varying voltage to this electrode \cite{RN503,RN504}. The electrostatic force gradient stemming from the electrode induces alterations in the spring constant.
    For a single NV center, the ground-state energy level structure is illustrated in Fig.~\ref{Fig:S8}(b). The ground triplet states are $\left| {{m}_{s}}=0,\pm 1 \right\rangle$.
    We applied a homogeneous static magnetic field ${{B}_{static}}$ to remove the degenerate states $\left| {{m}_{s}}=\pm 1 \right\rangle $ with the Zeeman splitting $\delta =2{{g}_{e}}{{\mu }_{B}}{{B}_{static}}$, where ${{g}_{e}}\simeq 2$ and ${{\mu }_{B}}=14MHz/mT$ are the $\text{NV}'\text{s}$ Lande factor and Bohr magneton, respectively.

    \begin{figure}[htbp]
    \includegraphics[angle=0,width=1\linewidth]{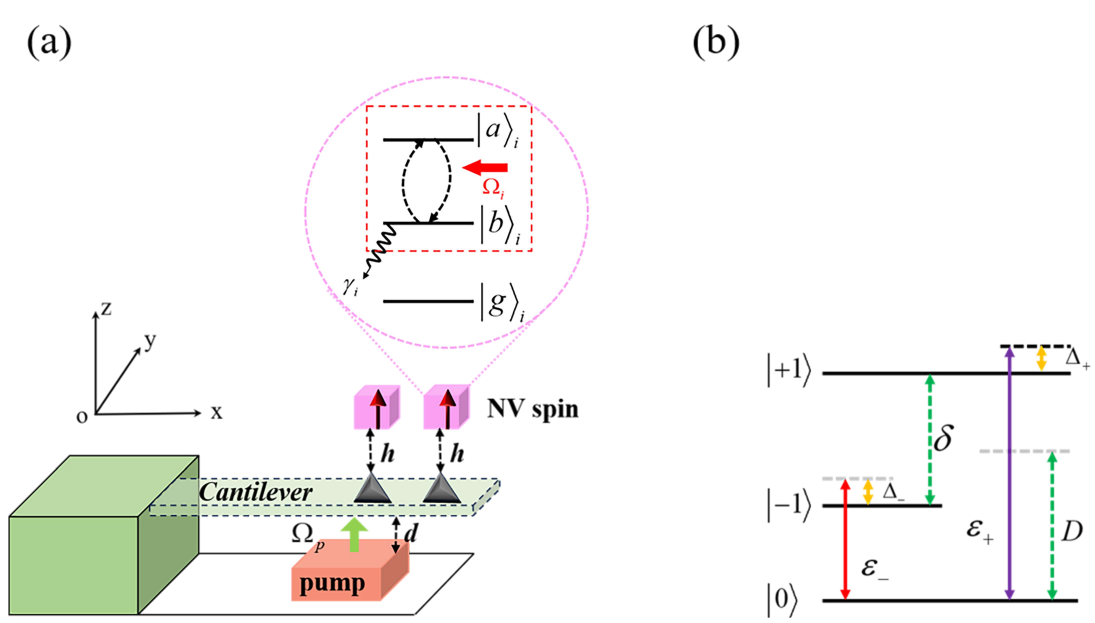}
    \caption{
    (a) Two magnet tips are placed at the end of the silicon cantilever. The spring constant of the cantilever is modified by the electric field from the capacitor plate. Two microwave fields polarized in the x direction (not shown in the picture) are applied to drive the NV centers between the state $\left| {{m}_{s}}=0 \right\rangle $ and the states $\left| {{m}_{s}}=\pm 1 \right\rangle $. (b) Level diagram of the driven NV center electronic ground state $\left| {{m}_{s}}=0,\pm 1 \right\rangle $.
    }
    \label{Fig:S8}
    \end{figure}
    
    By utilizing the two color microwave frequencies ${{\omega }_{\pm }}$, we can realize the transition between the states $\left| 0 \right\rangle $ and $\left| \pm 1 \right\rangle $. In the rotating frame with the microwave frequencies ${{\omega }_{\pm }}$, we obtain the Hamiltonian
    ${{H}_{NV}}=\sum\limits_{j=\pm 1}{-{{\Delta }_{j}}\left| j \right\rangle \left\langle  j \right|+\left( {{\varepsilon }_{j}}/2 \right)\left( \left| 0 \right\rangle \left\langle  j \right|+\left| j \right\rangle \left\langle  0 \right| \right)}$, where
    ${{\Delta }_{\pm }}\equiv \left| D-{{\omega }_{\pm }}\pm \delta /2 \right|$ and ${{\varepsilon }_{\pm }}\equiv {{g}_{e}}{{\mu }_{B}}B_{0}^{\pm }/\sqrt{2}$ with the microwave classical fields $B_{x}^{\pm }(t)=B_{0}^{\pm }\cos ({{\omega }_{\pm }}t+{{\phi }_{\pm }})$ polarized in the $x$ direction. In the following analysis, we take ${{\Delta }_{\pm }}=\Delta$ and ${{\varepsilon }_{\pm }}=\varepsilon$ for simplicity.
    The Hamiltonian for the nanomechanical resonator with a modulated spring is ${{H}_{mec}}=p_{z}^{2}/2M+\frac{1}{2}k(t){{z}^{2}}$, where ${{p}_{z}}$ and $z$ are the $\text{cantilever }\!\!'\!\!\text{ s}$ momentum and displacement operators, with effective mass $M$ and fundamental frequency ${{\omega }_{m}}$.
    Expressing the momentum operator ${{p}_{z}}$ and the displacement operator $z$ with the oscillator $a$ of the fundamental oscillating mode and the zero field fluctuation ${{z}_{zpf}}=\sqrt{\hbar /2M{{\omega }_{m}}}$, i.e., ${{p}_{z}}=-i\sqrt{M\hbar {{\omega }_{m}}/2}(a-{{a}^{\dagger }})$ and $z={{z}_{zpf}}(a+{{a}^{\dagger }})$.
    
    The Hamiltonian for describing the magnetic interaction between the NV spin and the cantilever vibrating mode can be written as ${{H}_{\operatorname{int}}}\equiv {{g}_{e}}{{\mu }_{B}}{{G}_{m}}z{{S}_{z}}$, with ${{G}_{m}}$ the magnetic field gradient.
    We assume that the two NV centers symmetrically placed on either side of the micromagnet along the direction of magnetization, and only consider the coupling between the oscillation in the $z$ direction and the NV center. Thus, the interaction between the NV centers can be limited in the $x$ direction \cite{PhysRevA.107.023722}.
    We switch to the dressed state basis $\left| b \right\rangle =1/\sqrt{2}\left( \left| +1 \right\rangle -\left| -1 \right\rangle  \right)$, $\left| g \right\rangle =\cos \theta \left| 0 \right\rangle -\sin \theta \left| h \right\rangle $, $\left| a \right\rangle =\cos \theta \left| h \right\rangle +\sin \theta \left| 0 \right\rangle $, with $\left| h \right\rangle =1/\sqrt{2}\left( \left| +1 \right\rangle +\left| -1 \right\rangle  \right)$ and $\tan (2\theta )=-\sqrt{2}\Omega /\Delta $. The lowest energy level $\left| g \right\rangle $ is a stable ground state, which can be treated as an effective continuum outside of the submanifolds $\left| a \right\rangle $ and $\left| b \right\rangle $.
    Furthermore, we assume the transition frequency between the dressed states $\left| a \right\rangle $ and $\left| b \right\rangle $ is almost the same as the oscillator frequency, i.e., ${{\omega }_{ab}}\sim{{\omega }_{m}}$.
    This system can be described as 
    \begin{align}
    {{H}_{tot}} &\simeq {{\delta }_{m}}{{a}^{\dagger }}a
    - \frac{{{\Omega }_{p}}}{2} \left( {{a}^{2}} + {{a}^{\dagger 2}} \right) \notag \\
    &\quad + \sum\limits_{i=1,2} \left[ {{g}_{i}} \left( {{a}^{\dagger }}\sigma _{i}^{-} + a\sigma _{i}^{\dagger } \right)
    + \frac{{{\delta }_{dg}}}{2}\sigma _{i}^{z} \right]
    \label{Eq:9}
    \end{align}
    where the coefficients are ${{\delta }_{m}}={{\omega }_{m}}-{{\omega }_{p}}$, ${{\delta }_{ab}}={{\omega }_{ab}}-{{\omega }_{p}}$, $g={{g}_{e}}{{\mu }_{B}}{{G}_{m}}{{z}_{zpf}}\sin \theta $, $\sigma _{i}^{z}\equiv {{\left| a \right\rangle }_{i}}\left\langle  a \right|-{{\left| b \right\rangle }_{i}}\left\langle  b \right|$, $\sigma _{i}^{+}\equiv {{\left| a \right\rangle }_{i}}\left\langle  b \right|$ and $\sigma _{i}^{-}\equiv {{\left| b \right\rangle }_{i}}\left\langle  a \right|$. Therefore, it is obvious that the single NV can be seen as the two-level emitter.
    By applying another classical field to drive the two spins (with amplitude $\Omega $), the driving Hamiltonian can be obtained as ${{H}_{dri}}=\Omega (\sigma _{1}^{x}\text{+}\sigma _{2}^{x})$. Thus, the total Hamiltonian of our system can be rewritten as
    \begin{equation}
    {{H}_{TO}}\simeq {{H}_{tot}}+{{H}_{dri}}.
    \label{Eq:10}
    \end{equation}

    Considering the Hamiltonian ~(\ref{Eq:10}), we can diagonalize the mechanical part of ${{H}_{TO}}$ by the unitary transmission ${{U}_{s}}(r)=\exp \left[ r({{a}^{2}}-{{a}^{\dagger 2}})/2 \right]$, where the squeezing parameter $r$ is defined via the relation $\tanh 2r={{\Omega }_{p}}/{{\delta }_{m}}$. Then, the total Hamiltonian ~(\ref{Eq:10}) in this squeezed frame can be obtained as 
    \begin{equation}
    \begin{aligned}
        H_{TO} &= H_{Rabi}^{S} + H_{sq} \\
        H_{Rabi}^{S} &= {{\Delta }_{m}}a_{s}^{\dagger }{{a}_{s}} + \sum\limits_{i=1,2} \left[ {{g}_{eff,i}}({{a}_{s}} + a_{s}^{\dagger })(\sigma _{i}^{\dagger } + \sigma _{i}^{-}) \right. \\
        &\quad + \left. \frac{{{\delta }_{dg}}}{2}\sigma _{i}^{z} \right] + {{H}_{dri}} \\
        H_{sq} &= \sum_{i=1,2} \frac{g_i e^{-r}}{2}(a_s - a_s^{\dagger})(\sigma_i^{\dagger} - \sigma_i^{-}).
    \end{aligned}
    \label{Eq:11}
    \end{equation}
    
    Here, ${{g}_{eff,i}}={{g}_{i}}{{e}^{r}}/2$. Because the item ${{e}^{-r}}$ decreases to zero as the squeezing parameter $r$ increases, the Hamiltonian ${{H}_{sq}}$ can be ignored.
    
\begin{figure}[htbp]
    \centering
    \includegraphics[angle=0,width=0.7\linewidth]{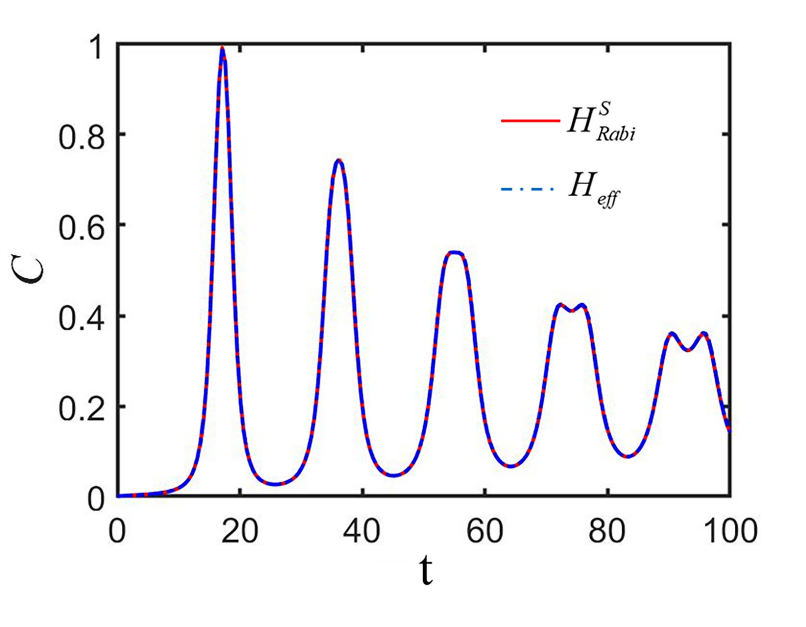}
    \caption{
    The concurrence evolution of the two spins system by respectively solving the two subequations of Eq.~(\ref{Eq:14}). The initial state of the system is $\left| 0 \right\rangle \left| aa \right\rangle $ and $\left| aa \right\rangle $ for Eq.~(\ref{Eq:14})(a) and Eq.~(\ref{Eq:14})(b), respectively. And the coefficients are ${{\delta }_{dg}}=0$, $\kappa _{a}^{S}=\gamma _{NV}^{i}=\gamma $, $\Omega/\gamma = 0.3$, $\xi/\gamma = 0.0006$, ${{\Delta }_{m}}/\gamma =40$ and ${{g}_{eff}}=\sqrt{\xi {{\Delta }_{m}}}$.
    }
    \label{Fig:S10}
    \end{figure}

    Considering ${{\delta }_{dg}}=0$ and ${{g}_{eff,1}}={{g}_{eff,2}}={{g}_{eff}}$, the Rabi Hamiltonian $H_{Rabi}^{S}$ can be reduced using the Schrieffer-Wolff transformation $H_{Rabi}^{eff}={{e}^{S}}H_{Rabi}^{S}{{e}^{-S}}$ where $S=\chi (a_{s}^{\dagger }-{{a}_{s}})(\sigma _{1}^{x}+\sigma _{2}^{x})$ and $\chi =\frac{{{g}_{eff}}}{{{\Delta }_{m}}}$. Noted that the parameter $\chi $ is much smaller than one, indicating that it satisfies the Lamb-Dicke condition $\chi \ll 1$. The effective Hamiltonian is given by 
    \begin{equation}
    H_{RT}^{eff}={{\Delta }_{m}}a_{s}^{\dagger }{{a}_{s}}-\xi {{\left( \sigma _{1}^{x}+\sigma _{2}^{x} \right)}^{2}}+\Omega (\sigma _{1}^{x}\text{+}\sigma _{2}^{x}).
    \label{Eq:12}
    \end{equation}
    where $\xi =\frac{g_{eff}^{2}}{{{\Delta }_{m}}}$. Owing to the operator ${{a}_{s}}$ is decoupled from the two-level emitter, we can only retain the second and third terms in Eq.~(\ref{Eq:11}). Especially, the second term is the typical Ising type interaction Hamiltonian ${{H}_{Ising}}=-\xi {{\left( \sigma _{1}^{x}+\sigma _{2}^{x} \right)}^{2}}$. Therefore, we rewrite the Hamiltonian (\ref{Eq:12}) as 
    \begin{equation}
    {H}_{eff}=-\xi {{(\sigma _{1}^{x}\text{+}\sigma _{2}^{x})}^{2}}+\Omega (\sigma _{1}^{x}\text{+}\sigma _{2}^{x}).
    \label{Eq:13}
    \end{equation}
    In this scenario, the effective spin-spin interaction of the two NVs is obtained, and the phonon is only virtually excited.
    Taking the effective dissipation $\kappa _{a}^{s}$ and the decaying rate of spin $\gamma _{NV}^{i}$ into consideration, we have the master equation as follow:
    \begin{subequations}
    \begin{align}
    & \frac{d\rho }{dt}=-i\left[ H_{Rabi}^{S},\rho  \right]+\kappa _{a}^{S}D[a]\rho +\sum\limits_{i=1,2}{\gamma _{NV}^{i}D}[{\sigma}_{i}^{-}]\rho,  \label{Eq:14a} \\
    & \frac{d\rho }{dt}=-i\left[ H_{eff},\rho  \right]+\sum\limits_{i=1,2}{\gamma _{NV}^{i}D}[{\sigma}_{i}^{-}]\rho. \label{Eq:14b}
    \end{align}
    \label{Eq:14}
    \end{subequations}
    where $D(O)\rho =O\rho {{O}^{\dagger }}-\frac{1}{2}\rho {{O}^{\dagger }}O-\frac{1}{2}{{O}^{\dagger }}O\rho $ is the Lindblad operator. Then we can make numerical simulations on the dynamical process according to Eq.~(\ref{Eq:14}). Figure~\ref{Fig:S10} displays the concurrence $C$ for the case of two NV spins varying with the evolution time by respectively solving the two subequations in Eq.~(\ref{Eq:14}). We can obviously find that our approximation is reasonable.
    
    \FloatBarrier

    \section{Non-Hermitian perturbation theory in Ising type interaction model}
    \label{Sec:A2}
    \setcounter{figure}{0} % Reset figure counter for each appendix section
    According to Eq.~(\ref{Eq:13}), the Hamiltonian for a single non-Hermitian qubit (or spin) reads as
    \begin{equation}
    \begin{aligned}
        & H_{0} = -\frac{i\gamma_{j}}{2}\left| b \right\rangle_{j}\left\langle b \right| + \Omega\left( \left| a \right\rangle_{j}\left\langle b \right| + \left| b \right\rangle_{j}\left\langle a \right| \right), \\
        & H_{Ising,j} = -\xi\left( \left| a \right\rangle_{j}\left\langle b \right| + \left| b \right\rangle_{j}\left\langle a \right| \right)^{2},
    \end{aligned}
    \label{Eq:A1}
    \end{equation}
    with eigenvalues
    \begin{equation}
    {{\lambda }_{j,\pm }}=\frac{1}{4}\left( -i\gamma -4\xi \pm \sqrt{16{{\Omega }^{2}}-{{\gamma }^{2}}} \right).
    \label{Eq:A2}
    \end{equation}
    The eigenstates in this single-qubit Hilbert space $H$ are given by
    \begin{equation}
    \begin{aligned}
        \left| \psi_{j,-} \right\rangle &= \frac{1}{4\sqrt{2}\Omega}
        \begin{pmatrix}
            i\gamma - \sqrt{16\Omega^2 - \gamma^2} \\
            4\Omega
        \end{pmatrix}, \\
        \left| \psi_{j,+} \right\rangle &= \frac{1}{4\sqrt{2}\Omega}
        \begin{pmatrix}
            i\gamma + \sqrt{16\Omega^2 - \gamma^2} \\
            4\Omega
        \end{pmatrix}.
    \end{aligned}
    \label{Eq:A3}
    \end{equation}
    It is easy to find that both the eigenvalues and eigenstates combine at the same critical point, i.e., EP, at $\eta \equiv \sqrt{16{{\Omega }^{2}}-{{\gamma }^{2}}}=0$. And the above states are normalized (i.e., $\left\langle  {{\psi }_{j,-(+)}} | {{\psi }_{j,-(+)}} \right\rangle =1$) but not orthogonal (i.e., $\left\langle  {{\psi }_{j,-}} | {{\psi }_{j,+}} \right\rangle \ne 0$).
    
    The non Hermiticity of $H$ is the main reason for the non-orthogonality, in order to better address the importance of non-Hermiticity, we extend the state basis to include states in the dual Hilbert space ${{H}^{*}}$. The Hamiltonian in the dual space ${{H}^{*}}$ is just the Hermitian conjugate of ${{H}_{j}}$ in Eq.~(\ref{Eq:A1}). i.e., ${{\overline{H}}_{j}}=H_{j}^{\dagger }$, with the eigenvalues ${{\overline{\lambda }}_{\pm }}=\frac{1}{4}\left( i\gamma -4\xi \pm \sqrt{16{{\Omega }^{2}}-{{\gamma }^{2}}} \right)$ and the corresponding eigenstates
    \begin{equation}
    \begin{aligned}
        \left| \overline{\psi}_{j,-} \right\rangle &= \frac{1}{4\sqrt{2}\Omega}
        \begin{pmatrix}
            -i\gamma - \sqrt{16\Omega^2 - \gamma^2} \\
            4\Omega
        \end{pmatrix}, \\
        \left| \overline{\psi}_{j,+} \right\rangle &= \frac{1}{4\sqrt{2}\Omega}
        \begin{pmatrix}
            -i\gamma + \sqrt{16\Omega^2 - \gamma^2} \\
            4\Omega
        \end{pmatrix}.
    \end{aligned}
    \label{Eq:A4}
    \end{equation}
    
    To describe the two coupled non-Hermitian qubit in the absence of Ising type interaction, we construct the eigenstates and eigenvalues of the Hamiltonian ${{H}_{0}}$ and its Hermitian conjugate ${{\overline{H}}_{0}}=H_{0}^{\dagger }$. Focused on the $\mathcal{P}\mathcal{T}$-symmetric region, i.e., $\Omega >{{\Omega }_{\text{EP4}}}$, that is, $\eta >0$.
    
    The first two eigenvalues of ${{H}_{0}}$ are given by
    \begin{equation}
    \begin{aligned}
        \lambda_{--} &= 2\lambda_{j,-} = \frac{1}{2}\left( -i\gamma - 4\xi - \eta \right), \\
        \lambda_{++} &= 2\lambda_{j,+} = \frac{1}{2}\left( -i\gamma - 4\xi + \eta \right).
    \end{aligned}
    \label{Eq:A5}
    \end{equation}
    and the corresponding eigenvalues in the two-qubit Hilbert space are the product states of the eigenstates of the two qubits:
    \begin{equation}
    \begin{aligned}
        \left| \psi_{--} \right\rangle &= \left| \psi_{1,-} \right\rangle \left| \psi_{2,-} \right\rangle = \frac{1}{32\Omega^2}
        \begin{pmatrix}
            -(\gamma + i\eta)^2 \\
            4\Omega(i\gamma - \eta) \\
            4\Omega(i\gamma - \eta) \\
            16\Omega^2
        \end{pmatrix}, \\
        \left| \psi_{++} \right\rangle &= \left| \psi_{1,+} \right\rangle \left| \psi_{2,+} \right\rangle = \frac{1}{32\Omega^2}
        \begin{pmatrix}
            (i\gamma + \eta)^2 \\
            4\Omega(i\gamma + \eta) \\
            4\Omega(i\gamma + \eta) \\
            16\Omega^2
        \end{pmatrix}.
    \end{aligned}
    \label{Eq:A6}
    \end{equation}
    As for the dual space, the eigenvalues of ${{\overline{H}}_{0}}$ become
    \begin{equation}
    \begin{aligned}
        \overline{\lambda}_{--} &= 2\overline{\lambda}_{j,-} = \frac{1}{2}\left( i\gamma - 4\xi - \eta \right), \\
        \overline{\lambda}_{++} &= 2\overline{\lambda}_{j,+} = \frac{1}{2}\left( i\gamma - 4\xi + \eta \right),
    \end{aligned}
    \label{Eq:A7}
    \end{equation}
    with the corresponding eigenstates 
    \begin{equation}
    \begin{aligned}
        \left| \overline{\psi}_{--} \right\rangle &= \left| \overline{\psi}_{1,-} \right\rangle \left| \overline{\psi}_{2,-} \right\rangle = \frac{1}{32\Omega^2}
        \begin{pmatrix}
            -(\gamma - i\eta)^2 \\
            -4\Omega(i\gamma + \eta) \\
            -4\Omega(i\gamma + \eta) \\
            16\Omega^2
        \end{pmatrix}, \\
        \left| \overline{\psi}_{++} \right\rangle &= \left| \overline{\psi}_{1,+} \right\rangle \left| \overline{\psi}_{2,+} \right\rangle = \frac{1}{32\Omega^2}
        \begin{pmatrix}
            -(\gamma + i\eta)^2 \\
            -4\Omega(-i\gamma + \eta) \\
            -4\Omega(-i\gamma + \eta) \\
            16\Omega^2
        \end{pmatrix}.
    \end{aligned}
    \label{Eq:A8}
    \end{equation}
    The above four eigenstates can be normalized via using biorthogonality
    \begin{equation}
    \begin{aligned}
        \left| \widetilde{\psi}_{--} \right\rangle &= \frac{\left| \psi_{--} \right\rangle}{\sqrt{\left\langle \overline{\psi}_{--} | \psi_{--} \right\rangle}} = \frac{1}{2\eta}
        \begin{pmatrix}
            -i\gamma + \eta \\
            -4\Omega \\
            -4\Omega \\
            i\gamma + \eta
        \end{pmatrix}, \\
        \left\langle \widetilde{\overline{\psi}}_{--} \right| &= \frac{\left\langle \overline{\psi}_{--} \right|}{\sqrt{\left\langle \overline{\psi}_{--} | \psi_{--} \right\rangle}} \\
        &= \frac{1}{2\eta}
        \begin{pmatrix}
            -i\gamma + \eta & -4\Omega & -4\Omega & i\gamma + \eta
        \end{pmatrix}, \\
        \left| \widetilde{\psi}_{++} \right\rangle &= \frac{\left| \psi_{++} \right\rangle}{\sqrt{\left\langle \overline{\psi}_{++} | \psi_{++} \right\rangle}} = \frac{1}{2\eta}
        \begin{pmatrix}
            i\gamma + \eta \\
            4\Omega \\
            4\Omega \\
            -i\gamma + \eta
        \end{pmatrix}, \\
        \left\langle \widetilde{\overline{\psi}}_{++} \right| &= \frac{\left\langle \overline{\psi}_{++} \right|}{\sqrt{\left\langle \overline{\psi}_{++} | \psi_{++} \right\rangle}} = \frac{1}{2\eta}
        \begin{pmatrix}
            i\gamma + \eta & 4\Omega & 4\Omega & -i\gamma + \eta
        \end{pmatrix}.
    \end{aligned}
    \label{Eq:A9}
    \end{equation}
    Here, $\left\langle  {{\overline{\psi }}_{j}} \right|(j=+,-)$ are the Hermitian conjugates of $\left| {{\overline{\psi }}_{j}} \right\rangle $ in the dual space.
    
    The other two eigenvalues of ${{H}_{0}}$ are degenerate and given by
    \begin{equation}
    {{\lambda }_{-+}}={{\lambda }_{+-}}={{\lambda }_{j,-}}+{{\lambda }_{j,+}}=-\frac{i\gamma }{2}-2\xi,
    \label{Eq:A10}
    \end{equation}
    with the corresponding eigenstates
    \begin{equation}
    \begin{aligned}
        \left| \psi_{-+} \right\rangle &= \left| \psi_{1,-} \right\rangle \left| \psi_{2,+} \right\rangle = \frac{1}{8\Omega}
        \begin{pmatrix}
            -4\Omega \\
            i\gamma - \eta \\
            i\gamma + \eta \\
            4\Omega
        \end{pmatrix}, \\
        \left| \psi_{+-} \right\rangle &= \left| \psi_{1,+} \right\rangle \left| \psi_{2,-} \right\rangle = \frac{1}{8\Omega}
        \begin{pmatrix}
            -4\Omega \\
            i\gamma + \eta \\
            i\gamma - \eta \\
            4\Omega
        \end{pmatrix}.
    \end{aligned}
    \label{Eq:A11}
    \end{equation}
    The eigenvalues and eigenstates of ${{\overline{H}}_{0}}$ in the dual space are given by,
    \begin{equation}
    {{\overline{\lambda }}_{-+}}={{\overline{\lambda }}_{+-}}={{\overline{\lambda }}_{j,-}}+{{\overline{\lambda }}_{j,+}}=\frac{i\gamma }{2}-2\xi.
    \label{Eq:A12}
    \end{equation}
    and the corresponding eigenstates are
    \begin{equation}
    \begin{aligned}
        \left| \overline{\psi}_{-+} \right\rangle &= \left| \overline{\psi}_{1,-} \right\rangle \left| \overline{\psi}_{2,+} \right\rangle = \frac{1}{8\Omega}
        \begin{pmatrix}
            -4\Omega \\
            -i\gamma - \eta \\
            -i\gamma + \eta \\
            4\Omega
        \end{pmatrix}, \\
        \left| \overline{\psi}_{+-} \right\rangle &= \left| \overline{\psi}_{1,+} \right\rangle \left| \overline{\psi}_{2,-} \right\rangle = \frac{1}{8\Omega}
        \begin{pmatrix}
            -4\Omega \\
            -i\gamma + \eta \\
            -i\gamma - \eta \\
            4\Omega
        \end{pmatrix}.
    \end{aligned}
    \label{Eq:A13}
    \end{equation}
    Similarly, we can build the normalized and unperturbed biorthogonal eigenstates in the degenerate subspace.
    \begin{equation}
    \begin{aligned}
        \left| \widetilde{\psi}_{-+} \right\rangle &= \frac{\left| \psi_{-+} \right\rangle}{\sqrt{\left\langle \overline{\psi}_{-+} | \psi_{-+} \right\rangle}} = \frac{1}{2\eta}
        \begin{pmatrix}
            -4\Omega \\
            i\gamma - \eta \\
            i\gamma + \eta \\
            4\Omega
        \end{pmatrix}, \\
        \left\langle \widetilde{\overline{\psi}}_{-+} \right| &= \frac{\left\langle \overline{\psi}_{-+} \right|}{\sqrt{\left\langle \overline{\psi}_{-+} | \psi_{-+} \right\rangle}} = \frac{1}{2\eta}
        \begin{pmatrix}
            -4\Omega & i\gamma - \eta & i\gamma + \eta & 4\Omega
        \end{pmatrix}, \\
        \left| \widetilde{\psi}_{+-} \right\rangle &= \frac{\left| \psi_{+-} \right\rangle}{\sqrt{\left\langle \overline{\psi}_{+-} | \psi_{+-} \right\rangle}} = \frac{1}{2\eta}
        \begin{pmatrix}
            -4\Omega \\
            i\gamma + \eta \\
            i\gamma - \eta \\
            4\Omega
        \end{pmatrix}, \\
        \left\langle \widetilde{\overline{\psi}}_{+-} \right| &= \frac{\left\langle \overline{\psi}_{+-} \right|}{\sqrt{\left\langle \overline{\psi}_{+-} | \psi_{+-} \right\rangle}} = \frac{1}{2\eta}
        \begin{pmatrix}
            -4\Omega & i\gamma + \eta & i\gamma - \eta & 4\Omega
        \end{pmatrix}.
    \end{aligned}
    \label{Eq:A14}
    \end{equation}
    
    \subsection{$\mathcal{P}\mathcal{T}$-symmetry-broken phase, $\eta <0$}
    \label{Sec:A3}
    For the $\mathcal{P}\mathcal{T}$-symmetry-broken phase, i.e., $\eta <0$, we can also present the conventionally normalized eigenvectors $\left| {{\psi }_{i}} \right\rangle $ in the broken phase, namely
    \begin{equation}
    \begin{aligned}
        & \left| {\psi}_{--} \right\rangle =\frac{1}{32{\Omega}^2}\begin{pmatrix}
       {(i\gamma +\eta)^2} \\
       4\Omega (i\gamma +\eta) \\
       4\Omega (i\gamma +\eta) \\
       16{\Omega}^2
    \end{pmatrix}, \\
        & \left| {\psi}_{++} \right\rangle =\frac{1}{32{\Omega}^2}\begin{pmatrix}
       {-(\gamma +i\eta)^2} \\
       4\Omega (i\gamma -\eta) \\
       4\Omega (i\gamma -\eta) \\
       16{\Omega}^2
    \end{pmatrix}, \\
        & \left| {\psi}_{-+} \right\rangle =\frac{1}{8\Omega}\begin{pmatrix}
       -4\Omega \\
       i\gamma +\eta \\
       i\gamma -\eta \\
       4\Omega
    \end{pmatrix},
         \left| {\psi}_{+-} \right\rangle =\frac{1}{8\Omega}\begin{pmatrix}
       -4\Omega \\
       i\gamma -\eta \\
       i\gamma +\eta \\
       4\Omega
    \end{pmatrix},
    \end{aligned}
    \label{Eq:B1}
    \end{equation}
    with corresponding eigenvalues calculated by ${{\lambda }_{--}}=2{{\lambda }_{j,-}}=\frac{1}{2}\left( -i\gamma -4\xi +\eta  \right),{{\lambda }_{++}}=2{{\lambda }_{j,+}}=\frac{1}{2}\left( -i\gamma -4\xi -\eta  \right)$ and ${{\lambda }_{-+}}={{\lambda }_{+-}}=-\frac{i\gamma }{2}-2\xi $, respectively.
    
    \subsection{First-order degenerate and non-degenerate perturbation theory with Ising type interaction}
    \label{Sec:A4}
    The perturbation matrix in the subspace is spanned by $\left\{ \left| {{\widetilde{\psi }}_{+-}} \right\rangle ,\left| {{\widetilde{\psi }}_{-+}} \right\rangle  \right\}$ and their adjoint states $\left\{ \left\langle  {{\widetilde{\overline{\psi }}}_{+-}} \right|,\left\langle  {{\widetilde{\overline{\psi }}}_{-+}} \right| \right\}$. This matrix can be obtained as
    \begin{equation}
    \begin{aligned}
        \left[ H_{\text{int}} \right] &=
        \begin{pmatrix}
            \left\langle \widetilde{\overline{\psi}}_{+-} \middle| H_{\text{int}} \middle| \widetilde{\psi}_{+-} \right\rangle & \left\langle \widetilde{\overline{\psi}}_{-+} \middle| H_{\text{int}} \middle| \widetilde{\psi}_{+-} \right\rangle \\
            \left\langle \widetilde{\overline{\psi}}_{-+} \middle| H_{\text{int}} \middle| \widetilde{\psi}_{+-} \right\rangle & \left\langle \widetilde{\overline{\psi}}_{-+} \middle| H_{\text{int}} \middle| \widetilde{\psi}_{-+} \right\rangle
        \end{pmatrix} \\
        &=
        \begin{pmatrix}
            -\frac{J\gamma^2}{\eta^2} & -\frac{J\gamma^2}{\eta^2} \\
            -\frac{J\gamma^2}{\eta^2} & -\frac{J\gamma^2}{\eta^2}
        \end{pmatrix}.
    \end{aligned}
    \label{Eq:C1}
    \end{equation}
    Here, $J=-2\xi $. Note that ${{H}_{\operatorname{int}}}$ is Hermitian and this submatrix is real with eigenvalues 0 and $-\frac{2J{{\gamma }^{2}}}{{{\eta }^{2}}}$. The eigenstates read as $\left( \left| {{\psi }_{-+}} \right\rangle \mp \left| {{\psi }_{+-}} \right\rangle  \right)/\sqrt{2}$, which means that we can choose a new basis for the degenerate subspace, specifically,
    \begin{equation}
    \begin{aligned}
        \left| \widetilde{\psi}_1 \right\rangle &= \frac{1}{\sqrt{2}} \left( \left| \widetilde{\psi}_{-+} \right\rangle - \left| \widetilde{\psi}_{+-} \right\rangle \right) 
        = \frac{1}{\sqrt{2}}
        \begin{pmatrix}
            0 \\
            -1 \\
            1 \\
            0
        \end{pmatrix}, \\[8pt]
        \left| \widetilde{\psi}_2 \right\rangle &= \frac{1}{\sqrt{2}} \left( \left| \widetilde{\psi}_{-+} \right\rangle + \left| \widetilde{\psi}_{+-} \right\rangle \right) 
        = \frac{1}{\sqrt{2}\eta}
        \begin{pmatrix}
            -4\Omega \\
            i\gamma \\
            i\gamma \\
            4\Omega
        \end{pmatrix}, \\[8pt]
        \left| \widetilde{\overline{\psi}}_1 \right\rangle &= \frac{1}{\sqrt{2}} \left( \left| \widetilde{\overline{\psi}}_{-+} \right\rangle - \left| \widetilde{\overline{\psi}}_{+-} \right\rangle \right) 
        = \frac{1}{\sqrt{2}}
        \begin{pmatrix}
            0 \\
            -1 \\
            1 \\
            0
        \end{pmatrix}, \\[8pt]
        \left| \widetilde{\overline{\psi}}_2 \right\rangle &= \frac{1}{\sqrt{2}} \left( \left| \widetilde{\overline{\psi}}_{-+} \right\rangle + \left| \widetilde{\overline{\psi}}_{+-} \right\rangle \right) 
        = \frac{1}{\sqrt{2}\eta}
        \begin{pmatrix}
            -4\Omega \\
            -i\gamma \\
            -i\gamma \\
            4\Omega
        \end{pmatrix}.
    \end{aligned}
    \label{Eq:C2}
    \end{equation}
    
    Note that this type of interaction Hamiltonian ${{H}_{\operatorname{int}}}$ is diagonal and the state $\left| {{\widetilde{\psi }}_{1}} \right\rangle $is equal to $\left| {{\widetilde{\overline{\psi }}}_{1}} \right\rangle $. The corresponding eigenvalues are given by
    \begin{equation}
    {{\lambda }_{1}}={{\lambda }_{-+}},{{\overline{\lambda }}_{1}}={{\overline{\lambda }}_{-+}},{{\lambda }_{2}}={{\lambda }_{+-}},{{\overline{\lambda }}_{2}}={{\overline{\lambda }}_{+-}}.
    \label{Eq:C3}
    \end{equation}
    It is easy to find that $\left| {{\widetilde{\psi }}_{1}} \right\rangle =\left| {{\widetilde{\overline{\psi }}}_{1}} \right\rangle $. We apply the first-order non degenerate perturbation theory to our system of two weakly coupled non-Hermitian spins. The perturbed eigenvalues are given by
    \begin{widetext}
    \begin{equation}
    \begin{aligned}
        \Lambda_1 &= \lambda_1 + 0 = -\frac{i\gamma}{2} - 2\xi, \quad
        \overline{\Lambda}_1 = \overline{\lambda}_1 + 0 = \frac{i\gamma}{2} - 2\xi, \\
        \Lambda_2 &= \lambda_2 - \frac{2J\gamma^2}{\eta^2} 
        = -\frac{i\gamma}{2} - 2\xi - \frac{2J\gamma^2}{\eta^2}, \quad
        \overline{\Lambda}_2 = \overline{\lambda}_2 - \frac{2J\gamma^2}{\eta^2} 
        = \frac{i\gamma}{2} - 2\xi - \frac{2J\gamma^2}{\eta^2}.
    \end{aligned}
    \label{Eq:C4}
    \end{equation}
    \end{widetext}
    
    We find that the given interaction Hamiltonian is already diagonal in the basis $\left\{ \left| {{\widetilde{\psi }}_{1}} \right\rangle ,\left| {{\widetilde{\psi }}_{2}} \right\rangle  \right\}$ of the Hilbert space and $\left\{ \left| {{\widetilde{\overline{\psi }}}_{1}} \right\rangle ,\left| {{\widetilde{\overline{\psi }}}_{2}} \right\rangle  \right\}$ of the dual space. Therefore, we need to consider the first-order unperturbed eigenstates of the non-degenerate subspace.
    
    \begin{widetext}
    \begin{equation}
        \left| {{\Psi }_{1}} \right\rangle =\left| {{\overline{\Psi }}_{1}} \right\rangle =\left| {{\widetilde{\psi }}_{1}} \right\rangle +\frac{\left\langle  {{\widetilde{\overline{\psi }}}_{--}} \right|{{H}_{\operatorname{int}}}\left| {{\widetilde{\psi }}_{1}} \right\rangle }{{{\lambda }_{1}}-{{\lambda }_{--}}}\left| {{\widetilde{\psi }}_{--}} \right\rangle +\frac{\left\langle  {{\widetilde{\overline{\psi }}}_{++}} \right|{{H}_{\operatorname{int}}}\left| {{\widetilde{\psi }}_{1}} \right\rangle }{{{\lambda }_{1}}-{{\lambda }_{++}}}\left| {{\widetilde{\psi }}_{++}} \right\rangle =\frac{1}{\sqrt{2}}{{\left( \begin{matrix}
       0 & -1 & 1 & 0  \\
    \end{matrix} \right)}^{\text{T}}},
    \label{Eq:C5}
    \end{equation}
    
    \begin{equation}
    \begin{aligned}
        \left| {{\Psi }_{2}} \right\rangle =\left| {{\widetilde{\psi }}_{2}} \right\rangle +\frac{\left\langle  {{\widetilde{\overline{\psi }}}_{--}} \right|{{H}_{\operatorname{int}}}\left| {{\widetilde{\psi }}_{2}} \right\rangle }{{{\lambda }_{2}}-{{\lambda }_{--}}}\left| {{\widetilde{\psi }}_{--}} \right\rangle +\frac{\left\langle  {{\widetilde{\overline{\psi }}}_{++}} \right|{{H}_{\operatorname{int}}}\left| {{\widetilde{\psi }}_{2}} \right\rangle }{{{\lambda }_{2}}-{{\lambda }_{++}}}\left| {{\widetilde{\psi }}_{++}} \right\rangle =\frac{1}{\sqrt{2}{{\eta }^{3}}}\left( \begin{matrix}
       -4\Omega {{\eta }^{2}}-i16\gamma J\Omega   \\
       i\gamma {{\eta }^{2}}  \\
       i\gamma {{\eta }^{2}}  \\
       4\Omega {{\eta }^{2}}-i16\gamma J\Omega   \\
    \end{matrix} \right),
    \end{aligned}
    \label{Eq:C6}
    \end{equation}
    
    \begin{equation}
    \begin{aligned}
        \left| {{\overline{\Psi }}_{2}} \right\rangle =\left| {{\widetilde{\overline{\psi }}}_{2}} \right\rangle +\frac{\left\langle  {{\widetilde{\psi }}_{--}} \right|{{H}_{\operatorname{int}}}\left| {{\widetilde{\overline{\psi }}}_{2}} \right\rangle }{{{\lambda }_{2}}-{{\lambda }_{--}}}\left| {{\widetilde{\overline{\psi }}}_{--}} \right\rangle +\frac{\left\langle  {{\widetilde{\psi }}_{++}} \right|{{H}_{\operatorname{int}}}\left| {{\widetilde{\overline{\psi }}}_{2}} \right\rangle }{{{\lambda }_{2}}-{{\lambda }_{++}}}\left| {{\widetilde{\overline{\psi }}}_{++}} \right\rangle =\frac{1}{\sqrt{2}{{\eta }^{3}}}\left( \begin{matrix}
       -4\Omega {{\eta }^{2}}+i16\gamma J\Omega   \\
       -i\gamma {{\eta }^{2}}  \\
       -i\gamma {{\eta }^{2}}  \\
       4\Omega {{\eta }^{2}}+i16\gamma J\Omega   \\
    \end{matrix} \right).
    \end{aligned}
    \label{Eq:C7}
    \end{equation}
    \end{widetext}
    
    The perturbed eigenvalues are given by
    \begin{widetext}
    \begin{equation}
    \begin{aligned}
        \left| \Psi_{++} \right\rangle &= \left| \widetilde{\psi}_{++} \right\rangle
        + \frac{\left\langle \widetilde{\overline{\psi}}_{--} \middle| H_{\text{int}} \middle| \widetilde{\psi}_{++} \right\rangle}{{\lambda_{++} - \lambda_{--}}} \left| \widetilde{\psi}_{--} \right\rangle
        + \frac{\left\langle \widetilde{\overline{\psi}}_{1} \middle| H_{\text{int}} \middle| \widetilde{\psi}_{++} \right\rangle}{{\lambda_{++} - \lambda_{1}}} \left| \widetilde{\psi}_{1} \right\rangle
        + \frac{\left\langle \widetilde{\overline{\psi}}_{2} \middle| H_{\text{int}} \middle| \widetilde{\psi}_{++} \right\rangle}{{\lambda_{++} - \lambda_{2}}} \left| \widetilde{\psi}_{2} \right\rangle \\
        &=\frac{1}{2\eta}
        \begin{pmatrix}
            i\gamma + \eta \\
            4\Omega \\
            4\Omega \\
            -i\gamma + \eta
        \end{pmatrix}
        - \frac{8J\Omega^2}{{\eta^4}}
        \begin{pmatrix}
            -i\gamma + \eta \\
            -4\Omega \\
            -4\Omega \\
            i\gamma + \eta
        \end{pmatrix}
        + \frac{8i\gamma J\Omega}{{\eta^4}}
        \begin{pmatrix}
            -4\Omega \\
            i\gamma \\
            i\gamma \\
            4\Omega
        \end{pmatrix},
    \end{aligned}
    \label{Eq:C8}
    \end{equation}
    
    \begin{equation}
    \begin{aligned}
        \left| \Psi_{--} \right\rangle &= \left| \widetilde{\psi}_{--} \right\rangle
        + \frac{\left\langle \widetilde{\overline{\psi}}_{++} \middle| H_{\text{int}} \middle| \widetilde{\psi}_{--} \right\rangle}{{\lambda_{--} - \lambda_{++}}} \left| \widetilde{\psi}_{++} \right\rangle
        + \frac{\left\langle \widetilde{\overline{\psi}}_{1} \middle| H_{\text{int}} \middle| \widetilde{\psi}_{--} \right\rangle}{{\lambda_{--} - \lambda_{1}}} \left| \widetilde{\psi}_{1} \right\rangle
        + \frac{\left\langle \widetilde{\overline{\psi}}_{2} \middle| H_{\text{int}} \middle| \widetilde{\psi}_{--} \right\rangle}{{\lambda_{--} - \lambda_{2}}} \left| \widetilde{\psi}_{2} \right\rangle \\
        &= \frac{1}{2\eta}
        \begin{pmatrix}
            -i\gamma + \eta \\
            -4\Omega \\
            -4\Omega \\
            i\gamma + \eta
        \end{pmatrix}
        + \frac{8J\Omega^2}{{\eta^4}}
        \begin{pmatrix}
            i\gamma + \eta \\
            4\Omega \\
            4\Omega \\
            -i\gamma + \eta
        \end{pmatrix}
        + \frac{8i\gamma J\Omega}{{\eta^4}}
        \begin{pmatrix}
            -4\Omega \\
            i\gamma \\
            i\gamma \\
            4\Omega
        \end{pmatrix}.
    \end{aligned}
    \label{Eq:C9}
    \end{equation}
    \end{widetext}
    
    The above perturbation theory gives the basis states, i.e., $\left| {{\psi }_{++}} \right\rangle ,\left| {{\psi }_{--}} \right\rangle ,\left| {{\psi }_{1}} \right\rangle ,\left| {{\psi }_{2}} \right\rangle $ in the Hilbert space $H$ as well as the corresponding adjoint states $\left| {{\overline{\psi }}_{++}} \right\rangle ,\left| {{\overline{\psi }}_{--}} \right\rangle ,\left| {{\overline{\psi }}_{1}} \right\rangle ,\left| {{\overline{\psi }}_{2}} \right\rangle $ in the dual space ${{H}^{*}}$. It is worth to note that the above basis states are both normalized
    \begin{equation}
    \left\langle  {{\overline{\psi }}_{1}} | {{\psi }_{1}} \right\rangle =1 and \left\langle  {{\overline{\psi }}_{i}} | {{\psi }_{i}} \right\rangle =1+O({{\xi }^{2}}),(i=++,--,2)
    \label{Eq:C10}
    \end{equation}
    and orthogonal, namely
    \begin{equation}
    \left\langle  {{\overline{\psi }}_{i}} | {{\psi }_{j}} \right\rangle =0+O({{\xi }^{2}}),(i,j=++,--,1,2)
    \label{Eq:C11}
    \end{equation}
    Moreover, they also compose a complete basis
    \begin{equation}
    \begin{aligned}
    \left| {{\psi }_{++}} \right\rangle \left\langle  {{\overline{\psi }}_{++}} \right| &+ \left| {{\psi }_{--}} \right\rangle \left\langle  {{\overline{\psi }}_{--}} \right| \\
    &+ \left| {{\psi }_{1}} \right\rangle \left\langle  {{\overline{\psi }}_{1}} \right| + \left| {{\psi }_{2}} \right\rangle \left\langle  {{\overline{\psi }}_{2}} \right| \\
    &= 1 + O({{\xi }^{2}}).
    \end{aligned}
    \label{Eq:C12}
    \end{equation}
    
    For the non-degenerate subspaces, the correction at the first order is determined by the expectation values of the interaction Hamiltonian using the unperturbed eigenstates
    \begin{widetext}
    \begin{equation}
    \begin{aligned}
        \Lambda_{--} = \lambda_{--} + \left\langle \widetilde{\overline{\psi}}_{--} \middle| H_{\text{int}} \middle| \widetilde{\psi}_{--} \right\rangle 
        = \frac{1}{2} \left( -i\gamma - 4\xi - \eta \right) + \frac{16J\Omega^2}{{\eta^2}}, \\
        \Lambda_{++} = \lambda_{++} + \left\langle \widetilde{\overline{\psi}}_{++} \middle| H_{\text{int}} \middle| \widetilde{\psi}_{++} \right\rangle 
        = \frac{1}{2} \left( -i\gamma - 4\xi + \eta \right) + \frac{16J\Omega^2}{{\eta^2}}.
    \end{aligned}
    \label{Eq:C13}
    \end{equation}
    \end{widetext}
    
    \subsection{Entanglement between two coupled non-Hermitian qubits}
    \label{Sec:A5}
    
    Given an initial state $\left| \psi (0) \right\rangle $, the time evolution of the state under an evolution operator $U={{e}^{-iHt}}$ can be obtained by
    \begin{widetext}
    \begin{equation}
    \begin{aligned}
        \left| \psi(t) \right\rangle &= U \left| \psi(0) \right\rangle \\
        &= \big\langle \overline{\Psi}_{++} \big| \psi(0) \big\rangle e^{-it\Lambda_{++}} \left| \Psi_{++} \right\rangle
        + \big\langle \overline{\Psi}_{--} \big| \psi(0) \big\rangle e^{-it\Lambda_{--}} \left| \Psi_{--} \right\rangle \\
        &\quad+ \big\langle \overline{\Psi}_{1} \big| \psi(0) \big\rangle e^{-it\Lambda_{1}} \left| \Psi_{1} \right\rangle
        + \big\langle \overline{\Psi}_{2} \big| \psi(0) \big\rangle e^{-it\Lambda_{2}} \left| \Psi_{2} \right\rangle \\
        &= A_1 e^{-it\Lambda_{++}} \left| \Psi_{++} \right\rangle 
        + A_2 e^{-it\Lambda_{--}} \left| \Psi_{--} \right\rangle 
        \quad+ A_3 e^{-it\Lambda_{1}} \left| \Psi_{1} \right\rangle 
        + A_4 e^{-it\Lambda_{2}} \left| \Psi_{2} \right\rangle,
    \end{aligned}
    \label{Eq:D1}
    \end{equation}
    \end{widetext}
    
    where
    \begin{equation}
    \begin{aligned}
        & A_1 = \frac{1}{2\eta}(i\gamma + \eta) - \frac{8J\Omega^2}{{\eta^4}}(-i\gamma + \eta) + \frac{8i\gamma J\Omega}{{\eta^4}}(-4\Omega) ,\\
        & A_2 = \frac{1}{2\eta}(-i\gamma + \eta) + \frac{8J\Omega^2}{{\eta^4}}(i\gamma + \eta) + \frac{8i\gamma J\Omega}{{\eta^4}}(-4\Omega) ,\\
        & A_3 = 0 ,\\
        & A_4 = \frac{1}{\sqrt{2}\eta^3}(-4\Omega\eta^2 - i16\gamma J\Omega).
    \end{aligned}
    \label{Eq:D2}
    \end{equation}
    By projecting this two-qubit pure state onto the four maximally entangled Bell states, i.e., $\left| \psi  \right\rangle =\sum\limits_{j}{{{c}_{j}}\left| {{e}_{j}} \right\rangle }$ with
    \begin{equation}
    \begin{aligned}
        & \left| e_1 \right\rangle = \frac{1}{\sqrt{2}}\left( \left| aa \right\rangle + \left| bb \right\rangle \right),
        & \left| e_2 \right\rangle = \frac{i}{\sqrt{2}}\left( \left| aa \right\rangle - \left| bb \right\rangle \right), \\
        & \left| e_3 \right\rangle = \frac{1}{\sqrt{2}}\left( \left| ab \right\rangle + \left| ba \right\rangle \right),
        & \left| e_4 \right\rangle = \frac{1}{\sqrt{2}}\left( \left| ab \right\rangle - \left| ba \right\rangle \right),
    \end{aligned}
    \label{Eq:D3}
    \end{equation}
    
    \begin{figure}[htbp]
    %\centering
    \includegraphics[angle=0,width=0.7\linewidth]{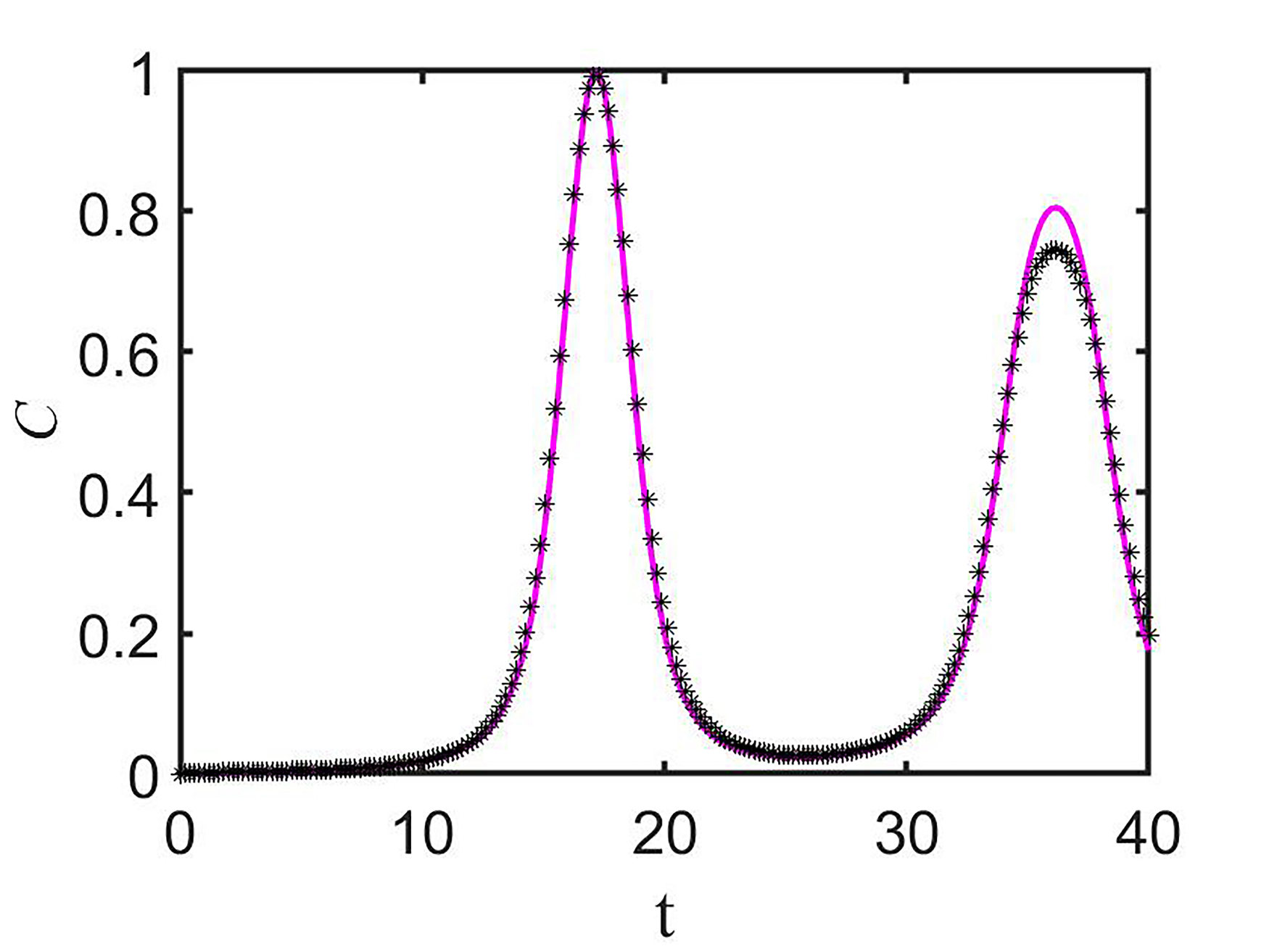}
    \caption{
    Comparison of the calculated concurrence evolution $C$ from the first-order non-Hermitian perturbation theory (the solid lines see Eq.~(\ref{Eq:D4})) and the fully numerical solution (asterisk on solid lines) at values in the $\mathcal{P}\mathcal{T}$-symmetric phase: $\Omega/\gamma = 0.3$ as an example. The initial state is $\left| \psi (0) \right\rangle =\left| aa \right\rangle $ and the coupling strength is $\xi/\gamma =0.0006$.
    }
    \label{Fig:S7}
    \end{figure}
    %\FloatBarrier
    
    The concurrence as an entanglement of the two coupled qubit can be written as
    \begin{equation}
    C=\frac{\left| \left\langle  {{\psi }^{*}} | \psi  \right\rangle  \right|}{{{\left| \left| \psi (t) \right\rangle  \right|}^{2}}}=\frac{2\left| \alpha \delta -\beta \varsigma  \right|}{{{\left| \alpha  \right|}^{2}}+{{\left| \beta  \right|}^{2}}+{{\left| \varsigma  \right|}^{2}}+{{\left| \delta  \right|}^{2}}}
    \label{Eq:D4}.
    \end{equation}
    The results show that the analytical simulation by utilizing the perturbation theory assuming the weak qubit coupling up to first order in the $\mathcal{P}\mathcal{T}$-symmetric phase. As an example, by choosing the weak Ising type interaction for $\xi/\gamma =0.0006$, the maximum entanglement ($C = 1$) will occur at the specific time ${{T}^{*}}=17.17$ for a given value of $\Omega/\gamma = 0.3$ in the PTS, as shown in Figure~\ref{Fig:S7}. Importantly, the analytical result shows good agreement with the numerical calculation.
    In particular, the maximum degree of entanglement can be obtained at a shorter timescale.
    
    \begin{figure}[htbp]
    %\centering
    \includegraphics[angle=0,width=1\linewidth]{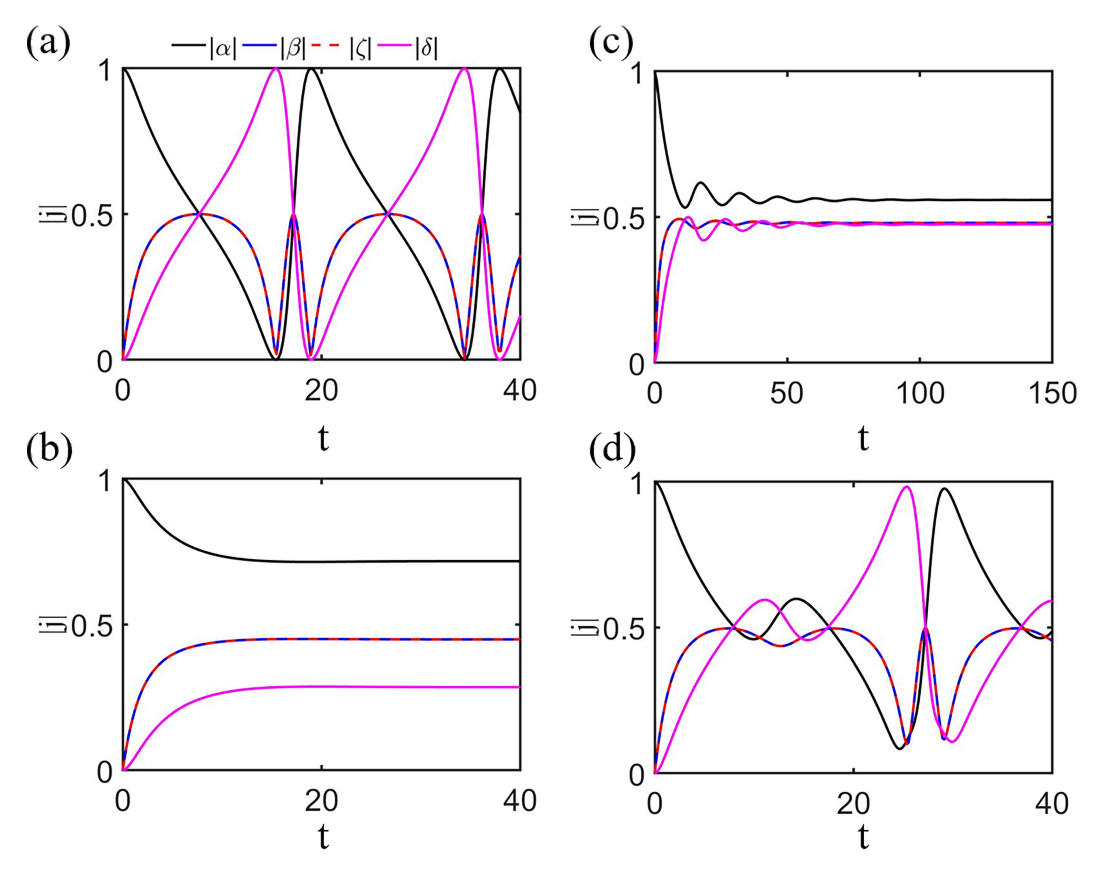}
    \caption{
      Time evolution of the modulus $\left| j \right| (j=\alpha ,\beta ,\varsigma ,\delta) $ for each of the four complex amplitudes of the
      two-qubit state $\left| \widetilde{\psi } \right\rangle =\alpha \left| aa \right\rangle +\beta \left| ab \right\rangle +\zeta \left| ba \right\rangle +\delta \left| bb \right\rangle $ for $\xi=0$ (a) and $\xi/\gamma=0.006$ (b-d). In the absence of Ising interaction, the evolution of $\left| j \right|$ is focused in the $\mathcal{P}\mathcal{T}$-symmetric phase. In the presence of Ising interaction, the evolution of $\left| j \right|$ has been shown in three different regions, i.e., in the $\mathcal{P}\mathcal{T}$-symmetry-broken phase (b), the mixed phase (c) and $\mathcal{P}\mathcal{T}$-symmetric phase (d), respectively. The given driving amplitude are chosen as $\Omega /\gamma =0.23$ (b), $\Omega /\gamma =0.28$ (c) and $\Omega /\gamma =0.3$ (d).
    }
    \label{Fig:S5}
    \end{figure}
    %\FloatBarrier
    
    According to Eq.~(\ref{Eq:D4}), the multitype entanglement dynamics (type I, II, III) can also be described in the aspect of the population of the basis state. Corresponding to the population at each basis state they exhibits different oscillatory behaviors in the three regions divided by the original EP4 and EP2 (as shown in Fig.~\ref{Fig:S5}), and different entanglement dynamics can be found in the three regions, accordingly. Compared with the population evolutions of the basis states $\left| bb \right\rangle $ and $\left| aa \right\rangle $ in the three different regions, we can find that the distorted Rabi-like oscillation of the population of $\left| bb \right\rangle $ and $\left| aa \right\rangle $ in the PTS, as the black and pink curves shown in Fig.~\ref{Fig:S5}(a) and ~\ref{Fig:S5}(d) regardless of whether there exists interaction.
    Moreover, the population evolution of the states $\left| ab \right\rangle $ and $\left| ba \right\rangle $ monotonically increase during the system evolving to a steady state in the PI [Fig.~\ref{Fig:S5}(b)] and the mixed phase [Fig.~\ref{Fig:S5}(c)]. Strikingly, the populations evolution of the basis states $\left| bb \right\rangle $ and $\left| aa \right\rangle $ are oscillations at short timescale and finally reach to a steady state in the mixed phase, which are different from that in PI. The different populations evolution corresponds to the distinctive entanglement dynamics in the mixed phase and PI.
    Notably, the populations of basis state are equal at specific time ${{T}^{*}}=27.2$, i.e., $\left| \alpha  \right|\sim\left| \beta  \right|\sim\left| \varsigma  \right|\sim\left| \delta  \right|$.
    
    \begin{figure}[htbp]
    \includegraphics[angle=0,width=1\linewidth]{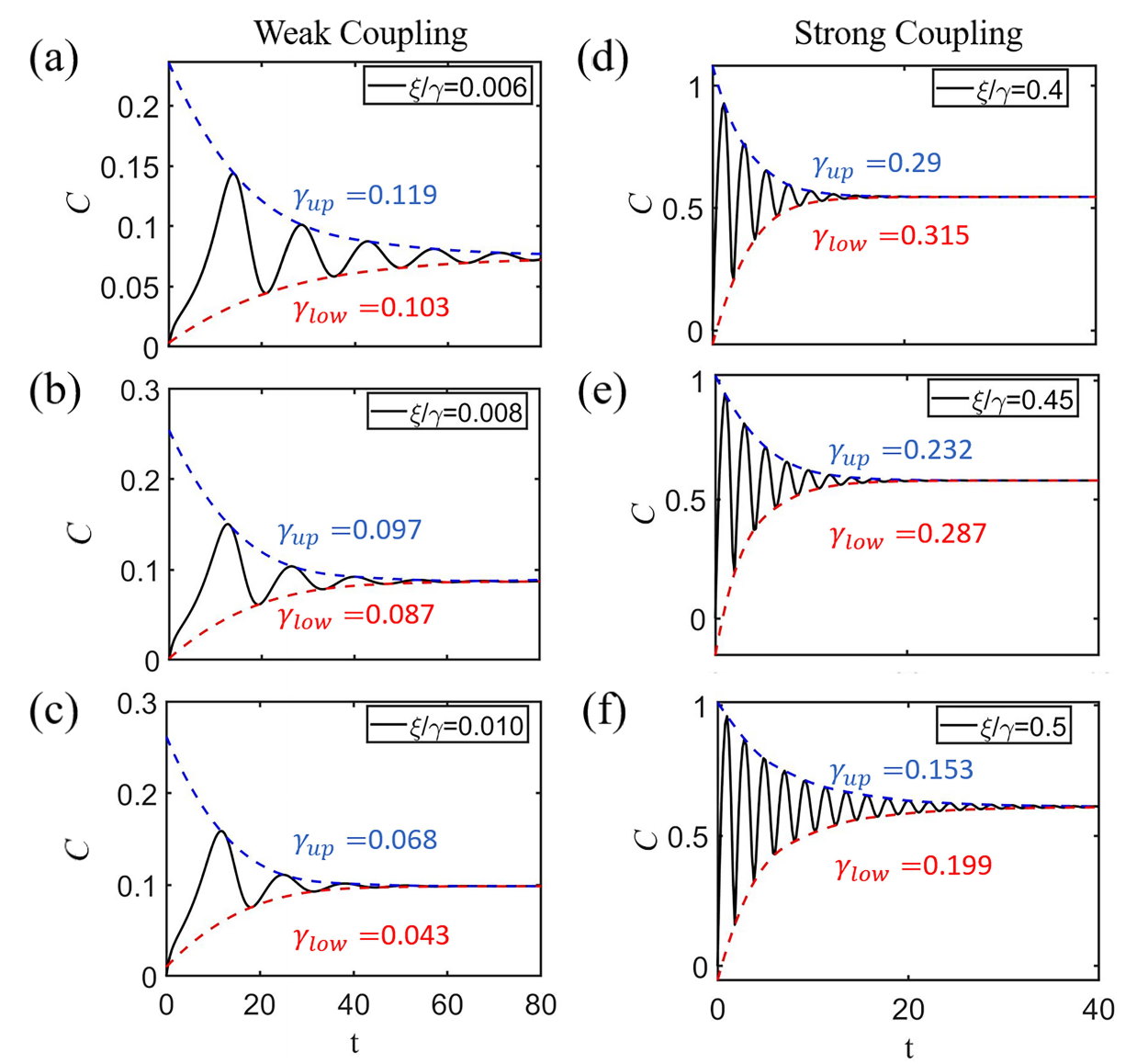}
    \caption{
    The decay rate of the upper (${{\gamma }_{up}}$) and lower (${{\gamma }_{low}}$) envelopes in the weak (a-c) and strong (d-f) coupling regimes. The dashed lines and solid lines are the fitting envelopes for the type III entanglement dynamics and concurrence evolution. 
    }
    \label{Fig:S12}
    \end{figure}
    
    Furthermore, we compare the decay rate of the fitting envelope in different coupling strength. We can find that both the decay rate of the upper (${{\gamma }_{up}}$) and lower envelopes (${{\gamma }_{low}}$) are decrease with the increase of the coupling strength. Additionally, in the weak coupling regime, the decay rates of the upper envelope are larger than that of the lower envelope, i.e., ${{\gamma }_{up}}>{{\gamma }_{low}}$ (see Fig.~\ref{Fig:S12} (a-c)). On the contrary, the decay rates of the upper envelope are smaller than that of the lower envelope, i.e., ${{\gamma }_{up}}<{{\gamma }_{low}}$, in the strong coupling regime (see Fig.~\ref{Fig:S12} (d-f)). 
    Additionally, we can only get the lower envelope fitting in the region of PI, and the relationship between the decay rate of upper and lower envelope fittings are not obvious. 
    
    \begin{acknowledgments}
        This work is supported by the National Natural Science Foundation of China (NSFC) (Grants No. 62075004, No. 11804018, No. U23A20481, No. 62275010, No. 12474353, and No. 12474354), Beijing Municipal Natural Science Foundation (Grants No. 4212051 and No. 1232027), and the Fundamental Research Funds for the Central Universities.
    \\
    \end{acknowledgments}
    \end{appendix}
    %\end{CJK}
    
    \bibliography{ref}
    
    \end{document}